\documentclass{aa}
\usepackage{epsf}
\usepackage{graphicx}

\newcommand{\taut}{\tau_\mathrm{T}}
\newcommand{\sigmat}{\sigma_\mathrm{T}}
\newcommand{\Ne}{n_\mathrm{e}}
\newcommand{\re}{r_\mathrm{e}}
\newcommand{\mpr}{m_\mathrm{p}}
\newcommand{\ms}{M_\odot}
\newcommand{\ledd}{L_\mathrm{Edd}}
\newcommand{\rc}{r_\mathrm{c}}

\newcommand{\lesssim}{\mathrel{\hbox{\rlap{\hbox{\lower4pt\hbox{$\sim$}}}\hbox{$<$}}}}
\newcommand{\gtrsim}{\mathrel{\hbox{\rlap{\hbox{\lower4pt\hbox{$\sim$}}}\hbox{$>$}}}}
\newcommand{\beq}{\begin{equation}}
\newcommand{\eeq}{\end{equation}}
\newcommand{\beqa}{\begin{eqnarray}}
\newcommand{\eeqa}{\end{eqnarray}}

\begin{document}

\title{Constraining the past X-ray luminosity of AGN in clusters of
galaxies: the role of resonant scattering}

\author{S.Yu. Sazonov$^{1,2}$, R.A. Sunyaev$^{1,2}$ \and C.K. Cramphorn$^{1}$} 

\offprints{sazonov@mpa-garching.mpg.de}

\institute{$^1$Max-Planck-Institut f\"ur Astrophysik,
              Karl-Schwarzschild-Str. 1, 
              85740 Garching bei M\"unchen, Germany\\
              $^2$Space Research Institute, Russian Academy of Sciences,
              Profsoyuznaya 84/32, 117997 Moscow, Russia\\}

\date{Received 12 April 2002; accepted 28 June 2002}

\titlerunning{The past X-ray luminosity of AGN}

\authorrunning{Sazonov et al.}

\maketitle

\begin{abstract}
Only a small fraction of galactic nuclei in the nearby universe are
luminous; most of them are currently dim. We investigate the
feasibility of constraining the X-ray luminosity in the recent past (up 
to $\sim 10^6$~years ago) of the nucleus of a cluster dominant galaxy
by measuring the contribution of scattered radiation from the central
source to the surface brightness of the intracluster gas dominated by
thermal emission. We show that resonance X-ray lines present an advantage
over the adjacent continuum, because the relative contribution
of the scattered component is typically larger in the line case by a
significant factor of 3--10. As an example, we estimate the level of
constraint that could be derived from future fine spectroscopic
observations on the past X-ray luminosity of the nearby M87 and
Cygnus~A active galaxies. For instance, scattered line radiation
should be detectable from the Virgo cluster if the X-ray luminosity of
M87 was a few times $10^{44}$~erg~s$^{-1}$, or $10^{-3}$ of its
Eddington luminosity, until a few times $10^5$~years ago. For
comparison, upper limits inferred from the available XMM-Newton and 
Chandra X-ray continuum data are typically 1--2 orders of magnitude weaker.

The same method can be applied to distant powerful quasars (at redshifts
$z\gtrsim 1$) if they have cluster-like gaseous coronae, as suggested
by Rosat and Chandra observations of active galaxies at $z\lesssim
1$. Their surface brightness profiles in the X-ray continuum above
$\sim 10$~keV $\gg kT/(1+z)$ (where $T$ is the gas temperature) should
be dominated by redshifted scattered radiation from the
quasar. Therefore, measurements with forthcoming mirror-optics hard
X-ray telescopes could give information on the lifetime of quasars and
parameters of the hot gas around them.


\end{abstract}

\section{Introduction}
\label{intro}

It is commonly believed that active galactic nuclei (AGN) are powered
by accretion onto a supermassive black hole. However, a detailed
description of this process is still missing. One of the principal
obstacles to making progress here is that we know very little from
observations about the variability of AGN on timescales longer than a
few tens of years.

Nuclear variability on the longest timescales up to the Hubble time
is expected to be governed by the supply of gas from galactic scales
to the central kpc. Such transport can be triggered by mergers and
interactions of galaxies (e.g. Hernquist \cite{hernquist89}). There
are reasons to believe that AGN can be strongly variable on timescales much
shorter than the Hubble time. For example, Ciotti \& Ostriker
(\cite{cioost97,cioost01}), in a follow-up to Binney \& Tabor
(\cite{bintab95}), have suggested a mechanism for 
elliptical galaxies in which an accreting central black hole heats the
ambient gas to the point at which the accretion stops, then fueling
resumes after the gas has cooled. In this model, strong outbursts
during which the AGN luminosity approaches the Eddington critical
value can occur every $10^8-10^9$~years. A similar outburst behaviour
but on shorter timescales of typically $10^6$~years can result from
thermal-viscous instability expected to operate in AGN accretion disks 
(Lin \& Shield \cite{linshi86}; Mineshige \&  Shields
\cite{minshi90}; Siemiginiowska, Czerny \& Kostyunin \cite{sieetal96};
Burderi, King \& Szuszkiewicz \cite{buretal98}). We may 
also mention tidal disruption of stars by supermassive black holes,
which can give rise to rare (every $10^{4}$--$10^{5}$~years per
galaxy), short (months to years) and bright (sub-Eddington) flares
(Rees \cite{rees88}).

Since we are unable to explore directly the long-term variability of
galactic nuclei, some indirect methods are in need. A feasible way
would be to observe radiation which was emitted in the past by a galactic
nucleus and later scattered toward us by the interstellar
medium. Since the characteristic size of a galaxy is a few tens kpc,
we could see the ``echo'' of previous AGN activity $\sim
10^5$~years after the central source turned off. This idea has been
put forward by Sunyaev, Markevitch \& Pavlinsky (\cite{sunetal93}) in
relation to our Galaxy. It was suggested that the diffuse,
hard X-ray emission observed today in the direction of giant molecular
complexes in the central  100~pc of the Milky Way is scattered radiation
emitted by Sgr A$^\ast$ in the past. This hypothesis has been confirmed by
the detection with the ASCA and Chandra satellites of diffuse emission
in the 6.4~keV fluoresence iron line from the Sgr B2 cloud (Koyama et
al. \cite{koyetal96}; Murakami et al. \cite{muretal01}). Cramphorn \& Sunyaev
(\cite{crasun02}) have recently elaborated on this problem and derived
upper limits on the X-ray luminosity of Sgr A$^\ast$ during the past
$10^5$~years.

It seems natural to extend the above approach to elliptical galaxies and
in particular to cluster dominant galaxies. Recent observations with the
Hubble Space Telescope and ground-based telescopes have confirmed previous
suggestions that 1) powerful low-redshift ($z< 0.5$) AGN are found in
luminous and preferentially early-type galaxies (Bahcall et
al. \cite{bahetal97}; Hooper, Impey \& Foltz \cite{hooetal97}; Boyce et
al. \cite{boyetal97}; McLure et al. \cite{mcletal99}; Schade, Boyle \&
Letawsky \cite{schetal00}) and 2) they are generally located in cluster
environments (McLure \& Dunlop \cite{mcldun01}). Since cluster
cores typically have sizes $\rc\sim 100$~kpc, we could hope to detect
radiation emitted by the nucleus (plus possibly by a jet) of a cluster
dominant galaxy until a few times $\rc/c\sim 10^6$~years ago and
scattered by the hot intracluster gas toward us. Up to $\sim 1$\% of
the radiation of the central source can be Thomson 
scattered by free electrons within the gas. 

Previous theoretical efforts in this direction have focused on the
prospects for detecting scattered (strongly polarized) AGN radiation at
radio and optical wavelengths (Sunyaev \cite{sunyaev82}; Gilfanov,
Sunyaev \& Churazov \cite{giletal87a}; Sholomitskii \& Yaskovich
\cite{shoyas90}; Wise \& Sarazin \cite{wissar90,wissar92}; Sarazin \&
Wise \cite{sarwis93}; Murphy \& Chernoff \cite{murche93}). However, it
would be particularly interesting to get information on the history
of the central source from observations of X-ray scattered
radiation, since the bulk of the AGN X-ray emission is believed to
originate within a few gravitational radii of the central black hole. 

One feasible way to detect the scattered X-rays against the thermal
emission of the hot gas would be to perform observations at $E\gg kT$, where
$E$ is the photon energy and $kT\sim 1$--10~keV is the gas
temperature. Indeed,  the AGN spectrum is a power-law, $F_\mathrm{AGN}\propto
E^{-\alpha}$ with $\alpha\sim 1$, while the plasma bremsstrahlung
spectrum is exponentially declining,
$F_\mathrm{gas}\propto\exp(-E/kT)$, at $E\gg kT$. Unfortunately, the angular
resolution and sensitivity of present-day hard X-ray telescopes are
not sufficient for mapping clusters of galaxies and detecting
scattered AGN radiation. However, the situation may improve
dramatically with the advent of grazing incidence X-ray telescopes
sensitive up to $\sim 40$~keV, such as those under 
consideration for the projected
Constellation-X\footnote{http://constellation.gsfc.nasa.gov/docs/main.html}
and XEUS\footnote{http://astro.esa.int/SA-general/Projects/XEUS/main/\\
xeus\_main.html}
missions. Furthermore, given the huge collecting area (several square
meters) of these planned telescopes in the standard X-ray band, it
should become possible to search for scattered X-rays 
from distant quasars (at redshifts $z\gtrsim 1$). Indeed, since the
radiation from a quasar gets redshifted on its way to us, a
significant gain in the ratio $F_\mathrm{AGN}/F_\mathrm{gas}\propto
[E(1+z)]^{-\alpha}\exp[E(1+z)/kT]$ could be achieved already at $E\sim
10$~keV. Such observations would permit one to constrain the lifetime of
quasars and the parameters of the hot gas around them. This
possibility is one of the issues discussed in the present paper
(particularly in \S\ref{cont}).

On the observational side, evidence has been accumulated over recent
years that AGN are often surrounded by gas atmospheres typical of clusters
of galaxies. Early ROSAT observations indicated the presence of luminous
extended X-ray emission around several powerful quasars with redshifts out
to $z=0.73$ (Crawford et al. \cite{craetal99}; Hardcastle \& Worrall
\cite{harwor99}). More recently, the Chandra observatory clearly
resolved extended hot gas around several relatively nearby AGN,
including 3C295 ($z=0.461$, Harris et al. \cite{haretal00}), 3C220.1
($z=0.62$, Worral et al. 2001\cite{woretal01}), 3C123 
($z=0.2177$, Hardcastle, Birkinshaw \& Worrall \cite{haretal01}),
Cyg~A ($z=0.0562$, Smith et al. \cite{smietal02}), and H1821+643
($z=0.297$, Fang et al. \cite{fanetal02}).

Chandra has also detected luminous extended (to radii $\sim
100$~kpc) X-ray emission surrounding two high-redshift powerful radio
galaxies 3C294 at $z=1.786$ (Fabian et al. \cite{fabetal01}) and
PKS~1138--262 at $z=2.156$ (Carilly et al. \cite{caretal02}). Assuming
for these objects that the emission of the central source is isotropic
and its X-ray luminosity has been the same in the past as it is now
($\sim$ a few $10^{45}$~erg s$^{-1}$), we are able to estimate (see
\S\ref{cont}) that the gas surface brightness should be  dominated by
scattered power-law radiation from the AGN already at $E\gtrsim 10$~keV.

In this paper we also discuss another possibility of detecting
scattered AGN X-ray radiation. We show (in \S\ref{lines_cont} and
\S\ref{line}) that at least for nearby AGN located in cluster dominant
galaxies or in isolated giant elliptical galaxies, one can gain
significantly by measuring the scattered surface brightness in
resonance X-ray lines rather than in the continuum. Such observations
would ideally require high ($\sim$~a few eV) spectral resolution in
order to separate the resonance lines from other lines. The
resolution needed together with high sensitivity can be achieved with
future high-energy astrophysics observatories such as
Astro-E2\footnote{http://www.astro.isas.ac.jp/astroe/index-e.html},
Constellation-X and 
XEUS. As an example, we estimate (in \S\ref{num_sim}) the level of
constraints it should be possible to derive on the past X-ray
luminosity of the nearby M87 and Cyg~A active galaxies. Yet better
constraints could be obtained by measuring the polarization of the
scattered X-ray resonance line radiation (see \S\ref{disc}; Sazonov et
al. \cite{sazetal02}).

\section{Advantage of scattered X-ray lines over the scattered continuum}
\label{lines_cont}

Consider an AGN surrounded by hot ($T\gtrsim 10^7$~K) gas. The
following emission components will contribute to the X-ray surface
brightness of the gas:
\begin{itemize}

\item
First, there is X-ray emission (line plus continuum) of the hot
plasma in which collisional ionization and excitation dominate. However,
a bright AGN may additionally photoionize the plasma. In our
calculations below we fully neglect this effect, although it can be
important when the AGN luminosity is very high (quasar-like), as will
be discussed in a separate paper (Sazonov \& Sunyaev, in preparation).

\item
Second, there is radiation emitted by the AGN and Thomson scattered by
free electrons in the gas. This scattered radiation contributes to the surface
brightness in the spectral continuum.

\item
Finally, AGN photons with energies falling within the cores of X-ray
lines of ions of heavy elements can resonantly scatter on these
ions, thus contributing to the surface brightness in the lines.
\end{itemize}

Consider an optically thin volume of gas with electron temperature $T$
and number density $\Ne$, exposed to an external spectral flux
$F_E(E)$ (measured in units of erg cm$^{-2}$~s$^{-1}$~keV$^{-1}$).

The plasma continuum spectral emissivity due to bremsstrahlung 
is given by (e.g. Zombeck \cite{zombeck90})
\beqa
\epsilon_{E,\mathrm{cont}}(E)=2.3\times 10^{-20} T^{-1/2}\exp(-E/kT)
\Ne^2 g_\mathrm{B}(T,E)
\nonumber\\
(\mathrm{erg}\,\mathrm{cm}^{-3}\,\mathrm{s}^{-1}\,\mathrm{keV}^{-1}),
\label{emis_cont}
\eeqa
where $g_\mathrm{B}$ is the corresponding Gaunt factor.

The energy-integrated emissivity in a resonance line by ions of type $z$ due to
electron collisional excitation of an electron from the ground level $i$ to an
excited level $k$ is given by (e.g. Zombeck \cite{zombeck90})
\beqa
\epsilon_\mathrm{line} &= & 2.7\times 10^{-15} T^{-1/2}\exp(-E_{i k}/kT)
\nonumber\\
&\times&\Ne n_z(T) f_{i k} g_{ik}(T)
\,\,(\mathrm{erg}\,\mathrm{cm}^{-3}\,\mathrm{s}^{-1}), 
\label{emis_line}
\eeqa
where $n_z$ is the ion number density, which depends on the gas
temperature. The remaining quantities appearing in equation
(\ref{emis_line}), namely $E_{i k}$, $f_{i k}$ and $g_{ik}$,
characterize the line itself, denoting its energy, oscillator strength
and excitation Gaunt factor, respectively.

The rate (per unit volume) of Thomson scattering of the external
radiation, integrated over the scattering angle $\theta$, is 
\beqa
\epsilon_{E,\mathrm{cont}}^\mathrm{scat}(E) &=& \frac{3}{8}
\int_{-1}^{1}d\cos\theta(1+\cos^2\theta)
\epsilon_{E,\theta,\mathrm{cont}}^\mathrm{scat}(E,\theta)
\nonumber\\
&=& F_E(E)\Ne\sigmat\,\,
(\mathrm{erg}\,\mathrm{cm}^{-3}\,\mathrm{s}^{-1}\,\mathrm{keV}^{-1})
\label{emis_thoms}
\eeqa
where $\sigmat=6.65\times 10^{-25}$ is the Thomson scattering cross
section. Note that we ignore throughout the Klein--Nishina correction
to the scattering cross section as well as the change in the photon
energy by scattering, which is justifiable if $E\ll mc^2$ and
$kT\ll mc^2$, where $mc^2=511$~keV is the electron rest energy.

The energy-integrated rate of resonant scattering (integrated over the
scattering angle) of the external radiation in the line $i\rightarrow
k$ by ions $z$ is
\beqa
\epsilon_\mathrm{line}^\mathrm{scat} &=& 4.1\times 10^{-18}\pi
\nonumber\\
&\times& F_E(E_{i k}) n_z(T)\re c f_{i k}\,\,
(\mathrm{erg}\,\mathrm{cm}^{-3}\,\mathrm{s}^{-1}),
\label{emis_res}
\eeqa
where $\re=2.82\times 10^{-13}$~cm is the classical electron radius,
$c$ is the speed of light, and the numerical coefficient accounts for
the unit transition from Hz to keV.

From equations (\ref{emis_cont}) and (\ref{emis_thoms}) we can find the
ratio of the Thomson scattering rate to the bremsstrahlung emissivity:
\beqa
\frac{\epsilon_{E,\mathrm{cont}}^\mathrm{scat}}
{\epsilon_{E,\mathrm{cont}}}(E) &=& 2.9\times 10^{-5}
\nonumber\\
&\times& F_E(E) T^{1/2}\exp(E/kT)\Ne^{-1}g^{-1}_\mathrm{B}(T,E).
\label{ratio_cont}
\eeqa

The ratio of the resonant scattering rate to the emissivity in the line
$i\rightarrow k$ is given by a similar expression, which follows from
equations (\ref{emis_line}) and (\ref{emis_res}),
\beqa
\frac{\epsilon_\mathrm{line}^\mathrm{scat}}{\epsilon_\mathrm{line}}
&=& 4.1\times 10^{-5}
\nonumber\\
&\times& F_E(E_{i k}) T^{1/2}\exp(E_{i k}/kT)\Ne^{-1}g^{-1}_{i k}(T).
\label{ratio_line}
\eeqa

We can now compare the ratios (\ref{ratio_line}) and
(\ref{ratio_cont}) for $E=E_{i k}$:
\beq
R\equiv\frac{\epsilon_\mathrm{line}^\mathrm{scat}}{\epsilon_\mathrm{line}}/
\frac{\epsilon_{E,\mathrm{cont}}^\mathrm{scat}}{\epsilon_{E,\mathrm{cont}}}=
1.4\frac{g_\mathrm{B}(T,E_{i k})}{g_{ik}(T)}.
\label{ratio}
\eeq

\begin{figure}
\centering
\includegraphics[width=\columnwidth]{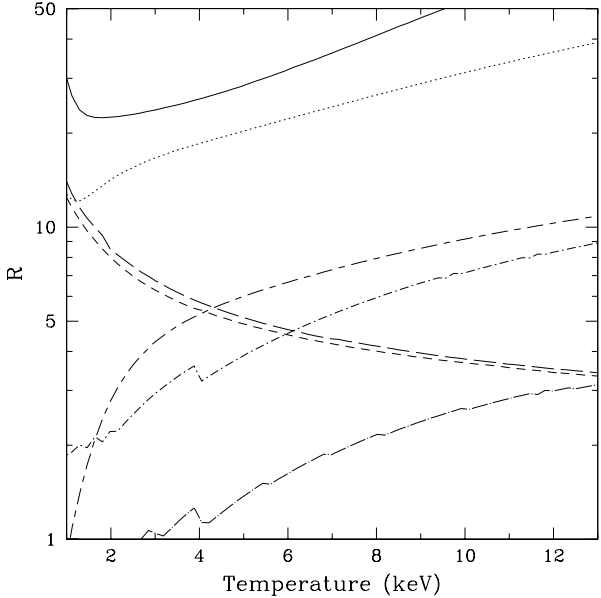}
\caption{Ratio $R$ defined by eq. [\ref{ratio}] as a function
of plasma temperature for various resonance X-ray lines: Fe~XXIII
L-shell line at 1.13~keV (solid), Fe~XXIV L-shell line at 1.17~keV
(dotted), Si~XIII K$\alpha$ line at 1.86~keV (short-dashed), S~XV
K$\alpha$ line at 2.46~keV (long-dashed), Fe~XXV K$\alpha$ line at
6.70~keV (dot-short-dashed), Fe~XXVI Ly$\alpha$ line at 6.97~keV
(dot-long-dashed), and Fe~XXV K$\beta$ line at 7.88~keV
(short-dash-long-dashed).
}
\label{r_temp}
\end{figure}

In the low-temperature limit ($kT\ll E_{ik}$), the excitation Gaunt
factor $g_{ik}$ ranges between 0.1 and 0.25 for most interesting lines;
it slowly increases with temperature and reaches 0.2--0.3 at 
$kT\sim E_{ik}$ (e.g. Mewe, Gronenschild \& van der Oord \cite{mewetal85}). The
bremsstrahlung Gaunt factor also is a weak function of temperature:
$g_\mathrm{B}=(kT/E)^{0.4}$ when $E\sim kT$. Summarizing these facts,
we may simply write
\beq
R\sim 6\,\,{\rm for}\, kT\sim E_{ik}.
\label{ratio_eff}
\eeq
This approximate relation is accurate to within a factor of 3 for all
resonance X-ray lines of interest to us. Fig.~\ref{r_temp} shows, for
several resonance lines, the dependence $R(T)$ computed using the
MEKAL plasma emission code in combination with other atomic data
sources (see \S\ref{num_sim} for details). In these computations the
contribution to $\epsilon_\mathrm{line}$ of unresolved (defined as
those with energies within one Doppler width of the resonance energy) satellite
lines was taken into account. This effectively leads to a smaller
value of $R$ compared to the definition (\ref{ratio}). However, this
reduction is only significant for the 6.70~keV and 6.97~keV K$\alpha$
lines or iron when $kT\ll E_{ik}$. We should note that the large
values ($R\sim 20$ at $kT\sim E_{ik}$) obtained for the iron L-shell
lines at 1.13~keV and 1.17~keV may be partly the result of the
different values for the oscillation strengths being used in the MEKAL
code with which we compute $\epsilon_\mathrm{line}$ and in the list of
resonance lines of Verner, Verner \& Ferland (\cite{veretal96}) which
we use to compute $\epsilon_\mathrm{line}^\mathrm{scat}$. We do not
attempt in this paper to correct for this and possibly other
inconsistencies caused by the simultaneous use of several sources of
atomic data. However, we estimate that some of our computational
results obtained below (in \S\ref{num_sim}) may contain a relative
systematic error of $\lesssim 2$.

We point out that $R$ is typically several times smaller for
intercombination lines than for resonance lines and $R\approx 0$ for
forbidden lines. For this reason we consider only resonance lines
throughout.

We have thus found that for a given resonance line with energy $E_{ik}$,
the relative contribution of scattered external radiation to the volume X-ray
emissivity of hot gas is typically larger by a significant factor of
the order of 3--10 than for the continuum at $E_{ik}$. 
We also know that for typical cluster temperatures $kT\sim$~a few keV
the combined surface brightness of the intracluster gas in the
resonance X-ray lines with $E_{ik}\sim kT$ is comparable to that in
the continuum. We therefore arrive at the conclusion that resonance X-ray
lines present significant advantage over the continuum for searches
of scattered AGN radiation in clusters of galaxies and elliptical
galaxies performed at $E\sim kT$. 

After we have introduced the $R$ factor in equation
(\ref{ratio}) and shown that this factor is only weakly dependent on
the resonance line and on gas temperature, we can continue our
treatment in parallel for continuum and line emission.

\section{AGN in the center of a beta-cluster}
\label{agn_beta}

\begin{figure}
\centering
\includegraphics[width=\columnwidth]{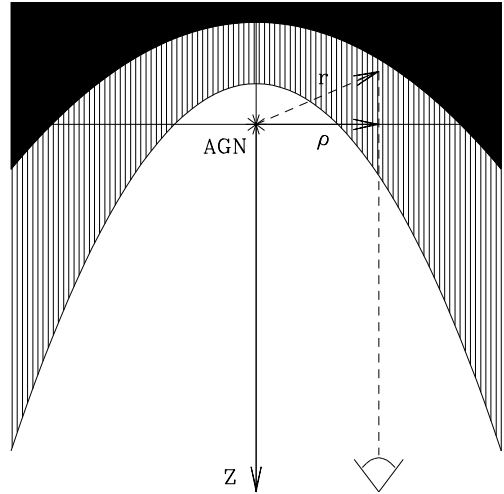}
\caption{Sketch of the model. The points with equal travel time from
the AGN to the observer lie on a paraboloid. In the switch-off case,
the thermal plasma located in the grey and black regions is scattering
the radiation emitted by the AGN. In the case of a flare of the AGN,
only the grey area is filled by photons and contributing to the
scattered emission. A possible path for a photon scattered into the
line of sight of the observer is marked by a dashed line. The system
is rotationally symmetric about the z-axis.}
\label{conrad}
\end{figure}

We shall now consider the following model (see Fig.~\ref{conrad}). A
supermassive black hole is located in the center of a cluster of
galaxies containing hot gas with a beta-law radial density profile
(Cavaliere \& Fusco-Femiano \cite{cavfus76}):
\beq
\Ne=n_0(1+r^2/\rc^2)^{-3\beta/2},
\label{dens_prof}
\eeq
where $n_0$ is the central electron number density and $\rc$ is the
core-radius. The gas temperature $T$ and element abundances
are constant over the cluster.

At a moment $t_\mathrm{on}=-\Delta$ (as measured by an earth bound
observer) the central source switches on (i.e. becomes an AGN) and at
a later moment $t_\mathrm{off}=0$ it switches off. The AGN emits
isotropically X-ray radiation having a power-law spectrum of photon
index $\gamma$, at a constant luminosity $L_X$ in the energy band 
$[E_1, E_2]$:
\beq
L_X=A\int_{E_1}^{E_2}
E^{-\gamma}E\,dE=A\frac{E_2^{2-\gamma}-E_1^{2-\gamma}}
{2-\gamma}.
\label{lum}
\eeq
The spectral flux at a given distance $r$ from the AGN is therefore
\beq
F_E=\frac{(2-\gamma)L_X}{4\pi(E_2^{2-\gamma}-E_1^{2-\gamma})r^2}E^{1-\gamma}.
\label{flux}
\eeq

Our goal is to find the surface brightness profile of the cluster,
including the contribution of scattered AGN radiation, at a given
moment of time $t>t_\mathrm{on}$.

\subsection{Approximation of isotropic scattering}
\label{isotropy}

We shall use throughout this paper the approximation of isotropic
scattering. This simplifies the treatment but requires explanation. We
know that in reality Thomson scattering occurs according to the
Rayleigh phase function $p(\cos{\theta})=3(1+\cos^2\theta)/8$. In the
case of resonant scattering, the phase function depends on the line
and generally can be represented as a weighted sum of the Rayleigh
phase function and the isotropic function $p(\cos{\theta})=1/2$
(see e.g. Chandrasekhar \cite{chandra50}). However, for our given
geometry of the problem the resulting brightness profiles depend only weakly
(typically by less than 10\%) on the phase function. For this reason,
the approximation of isotropic scattering is quite reasonable.

\subsection{Continuum radiation}
\label{cont}

We shall first consider the X-ray spectral continuum, for which the relevant
scattering mechanism is Thomson scattering. Since the Thomson optical
depths of clusters of galaxies and elliptical galaxies are small,
$\taut\lesssim 0.01$, we can use the single-scattering approximation
and make some analytic estimates.

The continuum (bremsstrahlung) surface brightness of a beta-cluster is
described, as a function of projected distance $\rho$ from the nucleus, by
the well-known formula, which can be obtained by integrating the plasma
emission along a given line of sight (i.e. along the coordinate $z$ -- see
Fig.~\ref{conrad}) using equations (\ref{emis_cont}) and
(\ref{dens_prof}):   
\beqa
B_{E,\mathrm{cont}} &\equiv& \int_{-\infty}^{\infty}
\frac{\epsilon_{E,\mathrm{cont}}(r)}{4\pi}\,dz
\nonumber\\
&=& 2.9\times 10^{-5}\frac{\Gamma(3\beta-1/2)}{\Gamma(3\beta)}
\left(\frac{n_0}{0.01\,\mathrm{cm}^{-3}}\right)^2\frac{\rc}
{100\,\mathrm{kpc}}
\nonumber\\
&\times& \left(\frac{kT}{1\,\mathrm{keV}}\right)^{-1/2}
(E/kT)^{-0.4}\exp(-E/kT)
\nonumber\\
&\times& (1+\rho^2/\rc^2)^{-3\beta+1/2}
\nonumber\\
&&
(\mathrm{erg}\,\mathrm{cm}^{-2}\,\mathrm{s}^{-1}\,\mathrm{sr^{-1}}
\,\mathrm{keV}^{-1}),
\label{bright_cont}
\eeqa
where $r=(\rho^2+z^2)^{1/2}$ and we have adopted
$g_\mathrm{B}(T,E)=(E/kT)^{-0.4}$.

The surface brightness of the Thomson scattered AGN radiation
is given by
\beq
B_{E,\mathrm{cont}}^\mathrm{scat} \equiv \int_{z_1}^{z_2}
\frac{\epsilon_{E,\mathrm{cont}}^\mathrm{scat}(r)}{4\pi}\,dz
=\frac{\sigma_\mathrm{T}}{4\pi}\int_{z_1}^{z_2}F_E(r) \Ne(r)\,dz,
\label{bright_scat}
\eeq
where we have used equation (\ref{emis_thoms}) and our assumption of
isotropic scattering.

The time-dependent integration limits $z_1(t,\rho)$ and $z_2(t,\rho)$
in equation (\ref{bright_scat}) are determined by the loci of
scattering sites giving a fixed time delay $\tilde{t}$. If
the distance between the emitter and the observer is much larger than
the characteristic size of the scattering cloud, which is true in our 
case, the constant-delay surface is a paraboloid with its focus at the
position of the source (see e.g. Sunyaev \& Churazov \cite{sunchu98} and
references therein to earlier work considering similar light-echo problems):
\beq
z(\tilde{t},\rho)=\frac{1}{2}\left(\frac{\rho^2}{c\tilde{t}}-c\tilde{t}\right).
\label{z_t}
\eeq
Therefore,
\beq 
z_1(t,\rho)=\frac{1}{2}\left[\frac{\rho^2}{c(t+\Delta)}
-c(t+\Delta)\right],
\label{z1}
\eeq
\beqa
z_2(t,\rho)=
\left\{
\begin{array}{ll}
(\rho^2/ct-ct)/2\,\,& \mathrm{if}\, 0<t\\
\infty \,\,& \mathrm{if}\,-\Delta<t<0.
\end{array}
\right.
\label{z2}
\eeqa

The integration limits in equation (\ref{bright_scat}) depend on the
outburst duration $\Delta$. We shall first consider two limiting cases
for which analytic treatment is possible. 

\subsubsection{Stationary case}
\label{stat}

Let the AGN be a persistent source. The limits of the integral 
$\int_{z_1}^{z_2}\epsilon_{E,\mathrm{cont}}^\mathrm{scat}\,dz$ are then
$z_1(t,\rho)=-\infty$ and $z_2(t,\rho)=\infty$, and it is possible, using
equation (\ref{dens_prof}) for $\Ne(r)$ and equation
(\ref{flux}) for $F_E$, to express the integral through hypergeometric
functions. In the particular interesting case of $\beta=2/3$, the
scattered surface brightness is
\beqa
B_{E,\mathrm{cont}}^\mathrm{scat}(\beta=2/3)=6.0\times 10^{-5}
\frac{L_X}{\ledd}\frac{M_\mathrm{BH}}{10^9\ms}
\frac{n_0}{0.01\,\mathrm{cm}^{-3}}
\nonumber\\
\times\frac{100\,\mathrm{kpc}}{\rc}
\frac{2-\gamma}{10^{2-\gamma}-1}
\left[\rc/\rho-(1+\rho^2/\rc^2)^{-1/2}\right]E^{1-\gamma}
\nonumber\\
(\mathrm{erg}\,\mathrm{cm}^{-2}\,\mathrm{s}^{-1}\,\mathrm{sr^{-1}}
\,\mathrm{keV}^{-1}),
\label{scat_stat_b23}
\eeqa
where $M_\mathrm{BH}$ is the mass of the central black hole,
$\ledd=1.4\times 10^{38}M_\mathrm{BH}/\ms$ is the corresponding
Eddington luminosity, $L_X$ is the AGN luminosity in the energy range
1--10~keV, and $E$ is measured in keV. 

The ratio of the scattered surface brightness to the
intrinsic surface brightness of the gas in the continuum is
\beqa
\frac{B_{E,\mathrm{cont}}^\mathrm{scat}}{B_{E,\mathrm{cont}}}
(\beta=2/3)=2.3\frac{L_X}{\ledd}\frac{M_\mathrm{BH}}{10^9\ms}
\frac{0.01\,\mathrm{cm}^{-3}}{n_0}
\nonumber\\
\times\left(\frac{100\,\mathrm{kpc}}{\rc}\right)^2
\left(\frac{kT}{1\,\mathrm{keV}}\right)^{0.1}\frac{2-\gamma}{10^{2-\gamma}-1}
\nonumber\\
\times (1+\rho^2/\rc^2)^{3/2}\left[
\rc/\rho-(1+\rho^2/\rc^2)^{-1/2}\right]
\nonumber\\
\times E^{1.4-\gamma}\exp(E/kT).
\label{ratio_stat_b23}
\eeqa

We can also write down asymptotic expressions
applicable in the limit of $\rho\gg\rc$ for arbitrary values of $\beta$:
\beqa
B_{E,\mathrm{cont}}^\mathrm{scat}(\rho\gg\rc)=3.2\times 10^{-5}
\frac{\Gamma(3\beta/2+1/2)}{\Gamma(3\beta/2+1)}\frac{L_X}{\ledd}
\frac{M_\mathrm{BH}}{10^9\ms}
\nonumber\\
\times\frac{n_0}{0.01\,\mathrm{cm}^{-3}}\frac{100\,\mathrm{kpc}}{\rc}
\frac{2-\gamma}{10^{2-\gamma}-1}
(\rho/\rc)^{-3\beta-1}E^{\gamma-1}
\nonumber\\
(\mathrm{erg}\,\mathrm{cm}^{-2}\,\mathrm{s}^{-1}\,\mathrm{sr^{-1}}
\,\mathrm{keV}^{-1});~~~
\label{scat_stat}
\eeqa
\beqa
\frac{B_{E,\mathrm{cont}}^\mathrm{scat}}{B_{E,\mathrm{cont}}}(\rho\gg\rc)= 
1.1\frac{\Gamma(3\beta/2+1/2)\Gamma(3\beta)}
{\Gamma(3\beta/2+1)\Gamma(3\beta-1/2)}\frac{L_X}{\ledd}
\nonumber\\
\times
\frac{M_\mathrm{BH}}{10^9\ms}\frac{0.01\,\mathrm{cm}^{-3}}{n_0}
\left(\frac{100\,\mathrm{kpc}}{\rc}\right)^2
\left(\frac{kT}{1\,\mathrm{keV}}\right)^{0.1}
\nonumber\\
\times
\frac{2-\gamma}{10^{2-\gamma}-1}
(\rho/\rc)^{3\beta-2}E^{1.4-\gamma}\exp(E/kT).
\label{ratio_stat}
\eeqa

\begin{figure}
\centering
\includegraphics[width=\columnwidth]{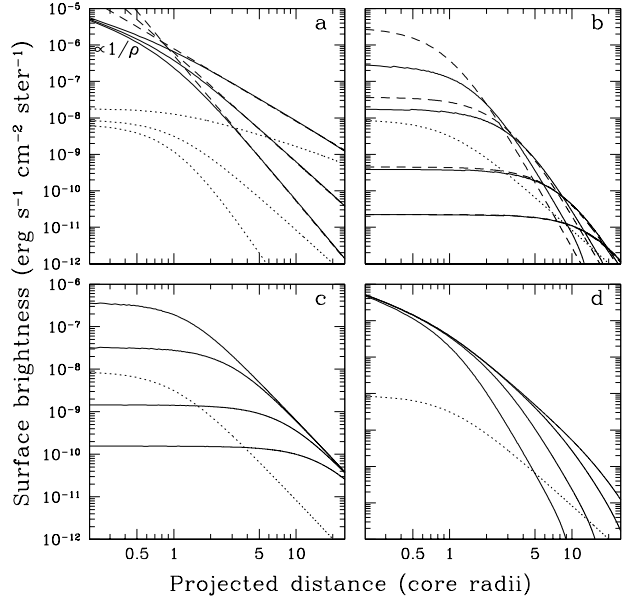}
\caption{{\bf a)} Surface brightness profiles of scattered AGN radiation (solid
lines) and plasma thermal emission (dotted lines) in the X-ray
continuum at $E=20$~keV, for a model in which a central $10^9\ms$
black hole is a persistent X-ray source with $L_X=\ledd$ and
$\gamma=2$. The host cluster has the following parameters:
$\rc=100$~kpc, $n_0=0.01$~cm$^{-3}$, $kT=3$~keV, and (from top to
bottom) $\beta=1/3$, $2/3$ and $1$, and is local ($z\ll 1$). The
dashed lines are $\rho\gg\rc$ asymptotes given by 
eq. [\ref{scat_stat}], applicable at $\rho\gtrsim 2\rc$. In the opposite limit 
of $\rho\ll 1$ the scattered surface brightness is proportional to
$\rho^{-1}$ and is independent of $\beta$, as indicated in 
the upper left corner of the figure. {\bf b)} Surface brightness profiles
of scattered AGN radiation (solid lines) at $E=20$~keV, measured at
different times ($t=\rc/c$, $3\rc/c$, $9\rc/c$ and 
$19\rc/c$ -- from top to bottom) after the end of a short
($\Delta=\rc/c$) flare with $L_X=\ledd$. For the same values of the
parameters as in (a) and $\beta=2/3$. The dashed lines are $\rho$,
$ct\gg\rc$ asymptotes given by eq. [\ref{scat_short}]. The dotted line
is the surface brightness profile of plasma thermal emission. {\bf c)} Same
as (b), but for the case where the central source was on ($L_X=\ledd$) until
$t_\mathrm{off}=0$ (profiles order from top to bottom). {\bf d)} Same as
(c), but for the case where the central AGN switched on at $t=0$
(profiles order from bottom to top).}
\label{cont_misc}
\end{figure}

It can also be shown that within the cluster core,
$B_{E,\mathrm{cont}}^\mathrm{scat}\propto \rho^{-1}$ and
$B_{E,\mathrm{cont}}^\mathrm{scat}/B_{E,\mathrm{cont}} \propto
\rho^{-1}$ for any $\beta$. This can be directly verified for
$\beta=2/3$ using equation (\ref{ratio_stat_b23}). 

\begin{figure}
\centering
\includegraphics[width=\columnwidth]{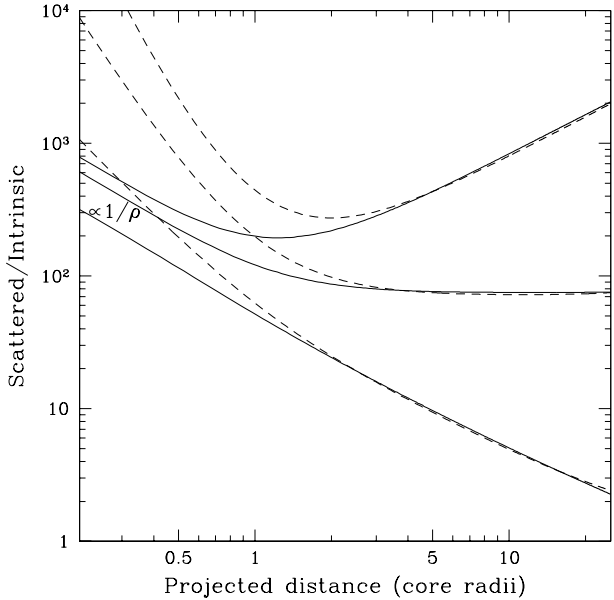}
\caption{Radial profiles of the ratio of the surface brightness of
scattered AGN radiation to the intrinsic brightness of the beta-cluster gas,
for the same model as in Fig.~\ref{cont_misc}a. The solid lines
correspond (from bottom to top) to $\beta=1/3$, $2/3$ and $1$. The
dashed lines are the corresponding $\rho\gg\rc$ asymptotes given by 
eq. [\ref{ratio_stat}]. 
} 
\label{cont_beta_ratio}
\end{figure}

Fig.~\ref{cont_misc}a demonstrates the dependence of the surface
brightness profile of a cluster with a central AGN on the
$\beta$ parameter. The scattered surface brightness profiles were derived by
numerical integration of equation
(\ref{bright_scat}). Fig.~\ref{cont_beta_ratio} shows the  
corresponding plots for the ratio of the scattered surface brightness to the 
intrinsic surface brightness of the intracluster gas. We see both from
Fig.~\ref{cont_beta_ratio} and equation (\ref{ratio_stat}) that
outside the core region (at $\rho\gtrsim\rc$), the relative
contribution of scattered AGN radiation to the surface brightness
increases with $\rho$ when $\beta>2/3$.

The $\beta=1/3$ case presented in Figs.~\ref{cont_misc}a and
\ref{cont_beta_ratio} corresponds to cluster cooling 
flows. Typically, gas density grows $\propto r^{-1}$ toward the 
center of the cooling flow, which roughly corresponds to a beta-model with a
small $\rc$, large $n_0$ and $\beta\approx 1/3$. We then find from
equations (\ref{bright_cont}) and (\ref{scat_stat}) that
$B_{E,\mathrm{cont}}\propto\rho^{-1}$ and 
$B_{E,\mathrm{cont}}^\mathrm{scat}\propto \rho^{-2}$ (assuming a
constant temperature) for cooling flows. Interestingly, the surface
brightness profile of scattered AGN radiation within the core of a
beta-cluster (see above) is similar ($\propto\rho^{-1}$) to the intrinsic
brightness profile of an (isothermal) cooling flow. Of course, a
cluster core dominated by scattered emission from a central powerful
quasar will have a very different spectrum (power law) than a cooling flow.

The energy dependent factor $E^{1.4-\gamma}\exp(E/kT)$ appearing in equations
(\ref{ratio_stat_b23}) and (\ref{ratio_stat}) has a minimum (for
$\gamma>1.4$) at $E=(\gamma-1.4)kT$ and increases exponentially when $E\gg 
kT$. The origin of this dependence is obvious: we are comparing the
slowly declining (power-law) spectrum of the AGN emission with the
exponentially decaying efficiency of collisional radiative mechanisms
operating in the intracluster gas. It is clear that for given AGN
luminosity and parameters of the gas, a significant gain (in terms of
the relative contribution of scattered AGN radiation to 
the surface brightness) can be achieved by performing observations at
$E\gg kT$, like in our examples in Figs.~\ref{cont_misc} and 
\ref{cont_beta_ratio}. A cautionary note is necessary here. If one
wants to constrain the past luminosity of an AGN by mapping its host
elliptical galaxy in hard X-rays, a restriction will appear due to a
significant contribution to the measured surface brightness from the 
(unresolved) X-ray binaries of the galaxy. We estimate that in
practice it should be very difficult to obtain limits better than $L_X\sim
10^{-3}\ledd$. Of course, this complication is irrelevant for 
surface brightness measurements carried out over regions of a galaxy cluster
that are almost devoid of point X-ray sources, say at distances $\sim
100$~kpc from the nucleus of a central dominant galaxy.

\subsubsection{The case of a high-redshift AGN}
\label{high-z}

Furthermore, if a powerful quasar is observed at high redshift $z$,
the relevant factor will be 
$[E(1+z)]^{1.4-\gamma}\exp[E(1+z)/kT]$. Consider for example
the radio galaxy 3C294 located at $z=1.786$, for which a
surrounding intracluster medium has recently been revealed with
Chandra (Fabian et al. \cite{fabetal01}). Since the total number of counts
registered from the source in the 20~ksec Chandra observation was small,
$\sim 100$, of which only $\sim 30$ are from the central AGN, it is only
possible to make rough estimates in this case. Fabian et al. find
that the gas temperature is $\sim 5$~keV, the 2--10~keV rest-frame
luminosity of the hot gas is $\sim 2.5\times 10^{44}$~erg~s$^{-1}$,
the corresponding (isotropic) luminosity of the AGN is
$\sim 1.1\times 10^{45}$~erg~s$^{-1}$, the photon index for the AGN
emission $\gamma\sim 2$ and the radial Thomson depth of the cluster is
$\sim 0.004$.

We can then estimate that the ratio of the flux of scattered
AGN radiation to that of thermal bremsstrahlung emission from the
gas is only $\sim 1$\% at $E\lesssim 2$~keV, $\sim 3$\% at
5~keV, $\sim 30$\% at 10~keV, $\sim 1$ at 12~keV and $\sim 4$ at
15~keV. These estimates assume that the central source is
isotropic and its luminosity has been the same in the recent past
(over $\sim$ a few $10^5$~years), and are made for the total
flux from the whole observed X-ray halo (of $\sim 15$~arcsec, or
equivalently $\sim 100$~kpc radius). We thus see that it is desirable
to go to energies above 10~keV in searching for scattered AGN
radiation around the 3C294 galaxy. However, photon statistics
then becomes a limiting factor. We can roughly estimate that the XEUS
observatory with its effective area of $\sim 2$~m$^2$ at 10~keV and
angular resolution of a few arcsec should be able to detect only a total of
$\sim 10$ scattered AGN photons above 10~keV in a 100~ks
observation of 3C294. This, combined with much more accurate
measurements of the surface brightness of the intracluster gas in the
standard X-ray band, could be only about enough to put constraints on
the past history of the 3C294 central X-ray luminosity.

Another similar example is the radio galaxy PKS~1138$-$262 at
$z=2.2$, from which Chandra also has observed luminous extended X-ray
emission (Carilly et al. \cite{caretal02}). In this case, the gas
X-ray luminosity is similar to that of the 3C294 cluster (unfortunately, the 
gas temperature is unconstrained), but the inferred 2--10~keV
(rest-frame) luminosity of the central source is four times 
higher, $\sim 4\times 10^{45}$~erg~s$^{-1}$. Moreover, in this case
the gas emission is likely dominated by scattered AGN
radiation already at $E\gtrsim 5$~keV (due to the higher redshift and
luminosity), so that a few hundred scattered AGN photons could
be detected above 5~keV in a 100~ks XEUS-2 observation 
of PKS~1138$-$262 (taking into account the larger collecting area of
the telescope at lower energies). 

We note that photon statistics should be less of a problem for more
nearby AGN and quasars such as 3C~273 located at $z=0.158$. In fact,
it is possible, given the huge X-ray luminosity of 3C~273, $\sim
2\times 10^{46}$~erg s$^{-1}$ (e.g. Yaqoob \& Serlemitsos
\cite{yaqser00}), that a hot intracluster medium surrounding it, if
any, will be first observed in scattered rather its own X-ray emission.  

\subsubsection{Short AGN flare}
\label{flare}

Suppose now that an AGN experienced an outburst in the past, between
$t_\mathrm{on}=-\Delta$ and $t_\mathrm{off}=0$. The ``short'' here
means that the line-of-sight depth of the illuminated (see
Fig.~\ref{conrad}) zone of the cluster must be much smaller than the 
characteristic size at a given projection radius. Therefore, one of the
following conditions must be fulfilled: $\Delta\ll\rc/c$ if
$\rho\lesssim\rc$ or $\Delta\ll\rho/c$ if $\rho\gtrsim\rc$. Then, as
follows from equations (\ref{z1}) and (\ref{z2}), 
\beq
\delta z(t,\rho)\equiv z_2-z_1\approx \frac{c\Delta}{2}
(1+\rho^2/c^2t^2),
\eeq
and
$\int_{z_1}^{z_2}\epsilon_{E,\mathrm{cont}}^\mathrm{scat}(\rho,z)\,dz\approx
\epsilon_{E,\mathrm{cont}}^\mathrm{scat}(\rho,z_2)\delta z(\rho)$. We
then find in the limit of $\rho$, $ct\gg\rc$ 
\beqa
B_{E,\mathrm{cont}}^\mathrm{scat} &=& 
3.2\times 10^{-5}
\times 4^{3\beta/2}\frac{L_X}{\ledd}\frac{M_\mathrm{BH}}{10^9\ms}
\nonumber\\
&\times&\frac{n_0}{0.01\,\mathrm{cm}^{-3}}\frac{100\,\mathrm{kpc}}{\rc}
\frac{2-\gamma}{10^{2-\gamma}-1}
\nonumber\\
&\times&
(\Delta/t)(\rc/ct)^{3\beta+1}(1+\rho^2/c^2t^2)^{-3\beta-1}E^{1-\gamma}
\nonumber\\
&&(\mathrm{erg}\,\mathrm{cm}^{-2}\,\mathrm{s}^{-1}\,\mathrm{sr^{-1}}
\,\mathrm{keV}^{-1});
\label{scat_short}
\eeqa
\beqa
\frac{B_{E,\mathrm{cont}}^\mathrm{scat}}{B_{E,\mathrm{cont}}}
&=& 1.2\times 4^{3\beta/2}\frac{\Gamma(3\beta)}{\Gamma(3\beta-1/2)}
\frac{L_X}{\ledd}\frac{M_\mathrm{BH}}{10^9\ms}
\nonumber\\
&\times&
\frac{0.01\,\mathrm{cm}^{-3}}{n_0}\left(\frac{100\,\mathrm{kpc}}{\rc}\right)^2
\left(\frac{kT}{1\,\mathrm{keV}}\right)^{0.1}\frac{2-\gamma}{10^{2-\gamma}-1}
\nonumber\\
&\times&
(\Delta/t)(ct/\rc)^{3\beta-2}
(\rho/ct)^{6\beta-1}
(1+\rho^2/c^2t^2)^{-3\beta-1}
\nonumber\\
&\times& E^{1.4-\gamma}\exp(E/kT).
\label{ratio_short}
\eeqa

Comparing equations (\ref{ratio_short}) and (\ref{ratio_stat}), we see that
for a given luminosity $L_X$ the contribution of scattered AGN radiation
to the surface brightness at $\rho\sim ct$ is smaller by a factor of
$\Delta/t$ for the flare campared to the stationary
case. This happens because in the former case the scattered
radiation comes from a gas layer of depth $\delta z\sim c\Delta $
along the line of sight, whereas $\delta z\sim\rho\sim ct$ in the latter case. 

Fig.~\ref{cont_misc}b shows a sequence of scattered
surface brightness profiles that would be measured at different times
after a short AGN outburst. We see that distant regions of the
cluster reveal no scattered light early on after the flare but
become progressively brighter as time goes by and photons of the flare
propagate through the gas. The ratio
$B_{E,\mathrm{cont}}^\mathrm{scat}/B_{E,\mathrm{cont}}$ has 
a maximum at $\rho_\mathrm{max}\sim ct$, as indicated by 
equation (\ref{ratio_short}). For example, $\rho_\mathrm{max}=\rc$ when
$\beta=2/3$. Therefore, the zone of largest contribution of scattered AGN
radiation propagates outward at an apparent velocity  equal to the speed
of light. It is important for observations that this zone is broad ---
$\delta\rho_\mathrm{eff}\sim\rho_\mathrm{max}\sim ct$.

We also notice that the amplitude of the effect is proportional to the
product $L_X\Delta$, i.e. to the total energy emitted by the AGN
during the flare. Thus, for example, a $10^5$-year outburst
at $L_X=0.01\ledd$ would produce the same scattered surface brightness
profile (at $\rho$, $ct\gg\rc$) as a $10^{4}$-year flare at $L_X=0.1\ledd$.

\subsubsection{Switch-off and switch-on scenarios}
\label{off_on}

We next consider a scenario in which the central source was
persistently (over a time that is longer than the 
characteristic light travel time of the cluster) bright in the past
until it suddenly turned off. A sequence of resulting scattered
surface brightness profiles corresponding to different times
after the switch-off ($t>0$) is shown in Fig.~\ref{cont_misc}c. One
can see that these profiles differ markedly from those
corresponding to the flare case (see Fig.~\ref{cont_misc}b). In
particular, as time goes by the surface brightness decreases in the
central region, while it remains practically unchanged in more distant
parts of the cluster still for a long time $\sim\rho/c$ after the
switch-off. This gives rise to a broad ($\delta\rho\sim\rho$) maximum
of the scattered/intrinsic brightness ratio, which propagates outward
at an apparent speed of light. 

Another interesting situation takes place when the central source is
bright at the time of observation but it has turned on recently. This
``switch-on'' case may actually correspond to the observed
quasars, which may have been luminous for a time short compared to the
light travel time of the gas around them. This situation is
depicted in Fig.~\ref{cont_misc}d. One can see that the radial
distribution of scattered AGN radiation is effectively truncated at a
certain radius $\sim c(t-t_\mathrm{on})$. This gives the possibility
to estimate the time during which the quasar has been active.

\subsection{Resonance lines}
\label{line}

The results obtained above can be directly extended to the case of
resonance lines provided the intracluster gas is optically thin
($\tau\ll 1$) to resonant scattering. We shall consider this limit
below and then (in \S\ref{depth}) point out differences that may arise
when $\tau\gtrsim 1$.

In the absence of external illumination, the equivalent width of a
line would be
\beqa
EW_0 &=&
\int_{-\infty}^{\infty}\epsilon_\mathrm{line}\,dz/\int_{-\infty}^{\infty}
\epsilon_{E,\mathrm{cont}}(E_{ik})\,dz
\nonumber\\
&=& 1.2\times 10^5\frac{n_z(T)}{\Ne}f_{ik}
\frac{g_{ik}(T)}{g_\mathrm{B}(T,E_{ik})}\,\,(\mathrm{keV}),
\label{w0}
\eeqa
under our assumptions of constant temperature and element abundances.

Resonant scattering of the AGN photons leads to an increase of
$EW$. At the same time, Thomson scattering of AGN radiation tends to
enhance the continuum near the line and thus to decrease its equivalent 
width. As we know from \S\ref{lines_cont}, the first effect is a 
factor of $R\sim 6$ larger than the second, hence the net effect of
the scattering of AGN emission in the intracluster gas should be
increased equivalent widths of the resonance lines. We can then write:
\beqa
EW &=& \left(\int_{-\infty}^{\infty}\epsilon_\mathrm{line}\,dz+\int_{z_1}^{z_2}
\epsilon_\mathrm{line}^\mathrm{scat}\,dz\right)/
\nonumber\\
&&\left(\int_{-\infty}^{\infty}\epsilon_{E,\mathrm{cont}}\,dz
+\int_{z_1}^{z_2}\epsilon_{E,\mathrm{cont}}^\mathrm{scat}\,dz\right)
\nonumber\\
&=&EW_0\frac{1+R\int_{z_1}^{z_2}\epsilon_{E,\mathrm{cont}}^\mathrm{scat}\,dz/
\int_{-\infty}^{\infty}\epsilon_{E,\mathrm{cont}}\,dz}
{1+\int_{z_1}^{z_2}\epsilon_{E,\mathrm{cont}}^\mathrm{scat}\,dz/
\int_{-\infty}^{\infty}\epsilon_{E,\mathrm{cont}}\,dz}.
\label{w}
\eeqa
 
If the contribution of scattered AGN radiation to the
surface brightness of the cluster is small, the fractional increase
of the equivalent line width will be 
\beqa
\frac{\Delta EW}{EW_0}&\equiv&\frac{EW-EW_0}{EW_0}\approx
\nonumber\\
&=&\frac{R-1}{5}\frac{\int_{z_1}^{z_2}
\epsilon_{E,\mathrm{cont}}^\mathrm{scat}(E_{ik})\,dz}
{\int_{-\infty}^{\infty}\epsilon_{E,\mathrm{cont}}(E_{ik})\,dz}.
\label{w_ratio}
\eeqa

Using equation (\ref{w_ratio}), one can readily apply all
the results of \S\ref{cont} to optically thin resonance lines. 

It should be emphasized that equation (\ref{w_ratio}) is strictly
valid only in the case where both the plasma temperature (which affects the
ionization balance and plasma emissivity) and the element abundances are
constant along the line of sight corresponding to a given projected distance
$\rho$. Indeed, only when these conditions are met,
$\epsilon_\mathrm{line}(r)\propto\epsilon_\mathrm{cont}(r)$ and
$\epsilon_\mathrm{line}^\mathrm{scat}(r)\propto
\epsilon_\mathrm{cont}^\mathrm{scat}(r)$ -- see equations
(\ref{emis_cont})--(\ref{emis_res}). In the general case, a
coefficient somewhat different than $R-1$ will relate the ratios
$\Delta EW/EW_0$ and
$\int_{z_1}^{z_2}\epsilon_{E,\mathrm{cont}}^\mathrm{scat}\,dz/ 
\int_{-\infty}^{\infty}\epsilon_{E,\mathrm{cont}}\,dz$. 

We should also remind the reader that we ignore throughout the effect
of the AGN on the ionization balance ($n_z$) in the intracluster medium. The
irradiation of the gas by a very bright central source 
will first of all alter the equivalent width of the emission line
relative to the value predicted by equation (\ref{w0}); it will also
affect (however to a smaller degree, because both $\epsilon_\mathrm{line}$
and $\epsilon_\mathrm{line}^\mathrm{scat}$ are proportional to $n_z$)
the ratio $\Delta EW/EW_0$.

\subsubsection{Finite depth effects}
\label{depth}

The intracluster gas may be moderately optically thick in the stronger
X-ray lines ($\tau\gtrsim 1$, e.g. Gilfanov, Sunyaev \& Churazov
\cite{giletal87b}), so the analytic results obtained above will be
only approximately valid in that case.

The optical depth at the center of a resonance line along a given direction
$\vec{l}$ through the cluster is 
\beq
\tau_0=\int\sigma_0 n_z(r)\,dl.
\label{tau}
\eeq
At plasma temperatures typical of galaxy clusters ($kT\sim$~
1--10~keV), all interesting X-ray lines have nearly Doppler  
absorption profiles whose width is determined by the velocities of
thermal and turbulent motions, because the line natural width is
relatively small. Therefore, the cross section at line center is
\beq
\sigma_0=\frac{\sqrt{\pi}h\re c f_{i k}}{\Delta E_\mathrm{D}}
\label{sigma_0}
\eeq
($h$ is Planck's constant), with the Doppler width being 
\beqa
\Delta E_\mathrm{D} &=& E_{i k}\left(\frac{2kT}{A\mpr c^2}+
\frac{V_\mathrm{turb}^2}{c^2}\right)^{1/2}
\nonumber\\
&=& E_{ik}\left[\frac{2kT}{A\mpr c^2}(1+1.4AM^2)\right]^{1/2},
\label{dop}
\eeqa
where $A$ is the element atomic mass, $\mpr$ is the proton
mass, $V_\mathrm{turb}$ is the characteristic turbulent
velocity and $M$ is the corresponding Mach number. The role of the
additional line broadening due to turbulence becomes more important
with increasing $A$.

Taking into account turbulence line broadening is important for two
reasons. First, this phenomenon may cause some resonance lines, which
would otherwise have $\tau\gtrsim 1$, to become 
optically thin. Secondly, some satellite lines may fall 
within the core of the resonance line, thereby effectively reducing the
factor $R$ defined in equation (\ref{ratio}). As was noted in
\S\ref{lines_cont}, this effect is particularly important for the
K$\alpha$ lines of H-like and He-like iron at $kT\sim 1$~keV.

If $\tau_0$ is non-negligible, first of all
the intrinsic surface brightness profile of the intracluster gas
will be distorted due to diffusion of photons from the central region to
the outer parts of the cluster. The equivalent line width will be
somewhat smaller than predicted by equation (\ref{w0}) at
$\rho\lesssim\rc$, and larger at $\rho\gtrsim\rc$ (Gilfanov et
al. \cite{giletal87b}). Note that this effect takes place already at
$\tau_0<1$, when only single resonant scattering is important, and becomes more
pronounced when $\tau_0>1$.

When the optical depth becomes larger than unity, multiple resonant
scatterings come into play. In the stationary scenario, the
surface brightness profile of scattered AGN emission in a line with
$\tau_0\gtrsim 1$ will be flatter in the cluster core compared to a
$1/\rho$ profile in the $\tau_0\ll 1$ case. Also,
the total flux of AGN photons scattered toward us from the core
region will be smaller than calculated in the single-scattering
approximation by a factor $\sim\tau_0$. However, due to the rather similar
effects taking place for thermal gas emission (see above), the
net correction to the ratio $EW/EW_0$ is expected to be fairly small
for $\tau_0\gtrsim 1$.

Multiple resonant scatterings also delay the escape of AGN photons
from the cluster core. As a result, the time-dependent scattered 
surface brightness profiles arising in the outburst scenario will be
affected. 

\section{Numerical simulations}
\label{num_sim}

As a next step, we have performed numerical simulations for two actual
clusters. We used a combination of 1) a Monte-Carlo code to play the
diffusion of photons (both those emitted by the AGN and by the gas)
through multiple resonant scatterings and 2)  the MEKAL code (Kaastra
\cite{kaastra92}) as implemented  in the software package XSPEC v10
(Arnaud \cite{arnaud96}) to calculate plasma emissivity in
lines. The line energies and oscillator strengths were taken from the
list of strong resonance lines of Verner, Verner \& Ferland
(\cite{verfer96}) and the solar abundances of elements from Anders \&
Grevesse (\cite{andgre89}). The ionization fractions were calculated
using collisional ionization rate fits from Voronov (\cite{voronov97}),
radiative recombination rates from Verner \& Ferland (\cite{verfer96})
and dielectronic recombination rates from Aldrovandi \& Pequignot
(\cite{aldpeq73}), Shull \& van Steenberg (\cite{shuste82}) and Arnaud \&
Rothenflug (\cite{arnrot85}).

\subsection{M87/Virgo}
\label{m87}

One of the most promising targets in the sky for 
observations of the discussed effect is M87. This is a giant
elliptical galaxy located near the center of the nearest ($17$~Mpc --
$1$~arcmin corresponds to 5~kpc) rich cluster of galaxies Virgo. M87
hosts a central black hole of mass $3\times 10^9\ms$ (Macchetto et
al. \cite{macetal97}; Marconi et al. \cite{maretal97}), which is the
largest reliably measured black hole mass. The corresponding Eddington
luminosity $\ledd= 4\times 10^{47}$~erg/s. 

M87 demonstrates AGN activity. In particular, observations at
different energy bands reveal an unresolved nucleus and a one-sided
jet. The combined bolometric luminosity of the nucleus and jet is
$\sim 10^{42}$~erg s$^{-1}$ (e.g. Biretta, Stern \& Harris \cite{biretal91}),
or $\sim 10^{-6}$ of the Eddington luminosity. This is an amazingly
small value given the fact that the nucleus is located in the center
of a dense cooling flow. The average power input in the form of
relativistic plasma from the jet into a large-scale radio halo around
M87 is estimated (Owen, Eilek \& Kassim \cite{oweetal00}) to have been
$\sim 10^{44}$~erg s$^{-1}$ over the past $\sim 10^8$~years, which is
still a small fraction of the Eddington luminosity.   

Both the nucleus and the jet (whose brightest knot is at a distance of
$\sim 1$~kpc from the nucleus) have recently been observed in X-rays
with the XMM-Newton (B\"{o}hringer et al. \cite{bohetal01}) and
Chandra (Wilson \& Yang \cite{wilyan02}) satellites. The X-ray
spectrum is power-law with values for the photon index ranging between
2 and 2.9 for the nucleus and different knots in the jet. The inferred
X-ray (1--10~keV) luminosities of the nucleus and the jet are $\sim
3\times 10^{40}$ and $8\times 10^{40}$~erg, respectively
(B\"{o}hringer et al. \cite{bohetal01}). 

\begin{table}
\caption{The brightest X-ray and extreme UV resonance lines of the
M87/Virgo intracluster gas.
}
\begin{center}
\begin{tabular}{lccc}\hline\hline
Ion & Energy & Equivalent width & Optical depth \\ 
    & (keV) &(eV)               &               \\
\hline
Fe XXIV  & 0.049 & 1   & 3.0 \\
Fe XXIV  & 0.065 & 2   & 6.0 \\
Fe XXIII & 0.093 & 3   & 8.5 \\
O VIII   & 0.654 & 20  & 0.6 \\
Ne X     & 1.021 & 30  & 0.9 \\ 
Fe XXIII & 1.129 & 10  & 3.2 \\
Fe XXIV  & 1.166 & 30  & 2.7 \\
Si XIII  & 1.865 & 10  & 1.3 \\
Si XIV   & 2.005 & 60  & 1.3 \\
S  XV    & 2.461 & 20  & 1.3 \\
Ar XVII  & 3.140 & 20  & 0.4 \\
Fe XXV   & 6.700 & 740 & 1.2 \\
Fe XXV   & 7.881 & 110 & 0.2 \\
\hline
\end{tabular}
\end{center}
\label{m87_lines}
\end{table}

In our simulations, we modeled the distribution of intracluster gas
around M87 based on the results of recent analyses of the XMM 
observations by B\"{o}hringer et al. (\cite{bohetal01}), Finoguenov et
al. (\cite{finetal02}) and Matsushita et al. (\cite{matetal02}). We
assumed the distribution of gas to be spherically symmetric around M87
and restricted our consideration to the central region with a radius
of 250~kpc. The electron density radial profile was represented by a sum of two
beta-models (\ref{dens_prof}) with the following parameters: 
$n_0=0.13$~cm$^{-3}$, $\rc=1.7$~kpc, $\beta=0.42$ and
$n_0=0.011$~cm$^{-3}$, $\rc=22$~kpc, $\beta=0.47$. We adopted a
constant temperature $kT=1$~keV within the central 1~kpc,
the dependence $kT=1.5(r/5\,\mathrm{kpc})^{0.22}$~keV between 1 and
50~kpc, and a constant $kT=2.5$~keV at larger radii. The last value 
approximately corresponds to temperatures measured with ASCA
(Shibata et al. \cite{shietal01}) in different parts of the Virgo
cluster within $\sim 300$~kpc from M87. The element abundances were
taken to be constant at $r<10$~kpc, with $A(\mathrm{O})=0.4$,
$A(\mathrm{Si})=1.0$, $A(\mathrm{S})=1.0$, $A(\mathrm{Ar})=1.0$ and
$A(\mathrm{Fe})=0.8$ (in units of solar abundances), then gradually
declining to $A(\mathrm{O})=0.4$, $A(\mathrm{Si})=0.6$,
$A(\mathrm{S})=0.6$, $A(\mathrm{Ar})=0.6$ and $A(\mathrm{Fe})=0.4$ at
$r=40$~kpc, and constant thereafter.

Table~\ref{m87_lines} lists resonance X-ray and extreme UV lines that
we estimate to be the strongest emission lines of the M87 gas. All
of these lines have substantial optical depths from the center to the
boundary of the gas cloud, assuming negligible turbulence.

Using equation (\ref{ratio_stat}), we can estimate for the stationary
scenario the contribution of scattered AGN radiation to the surface
brightness of M87/Virgo in the X-ray continuum:
\beqa
\frac{B_{E,\mathrm{cont}}^\mathrm{scat}}{B_{E,\mathrm{cont}}}
&=& 4\times 10^2\frac{L_X}{\ledd}E^{-0.6}\exp(E/1.5\,\mathrm{keV})
\nonumber\\
&&\times \left(\frac{\rho}
{1.7\,\mathrm{kpc}}\right)^{-0.7},\,\,
1.7\,\mathrm{kpc} \ll\rho\ll 22\,\mathrm{kpc};
\label{ratio_m87_1}
\eeqa
\beqa
\frac{B_{E,\mathrm{cont}}^\mathrm{scat}}{B_{E,\mathrm{cont}}}
&=& 40\frac{L_X}{\ledd}E^{-0.6}\exp(E/2\,\mathrm{keV})
\nonumber\\
&&\times
\left(\frac{\rho}{22\,\mathrm{kpc}}\right)^{-0.6},\,\,
\rho\gg 22\,\mathrm{kpc}.
\label{ratio_m87_2}
\eeqa
Here we have assumed constant temperatures of $kT=1.5$~keV and
$kT=2$~keV for equations (\ref{ratio_m87_1}) and (\ref{ratio_m87_2}),
respectively, and $\gamma=2$.

The corresponding dependences for the equivalent width of a resonance line with
$\tau_0\ll 1$ can be found from equations (\ref{ratio_m87_1}) and
(\ref{ratio_m87_2}) as $\Delta
EW/EW_0=(R-1)B_{E,\mathrm{cont}}^\mathrm{scat}(E_{ik})/B_{E,\mathrm{cont}}
(E_{ik})$ (assuming constant element abundances for a given $\rho$).

Now we proceed to discussing the results of our computations, which
do not use the assumption of small optical depth.

\subsubsection{Switch-off scenario}
\label{m87_off}

\begin{figure}
\centering
\includegraphics[width=\columnwidth]{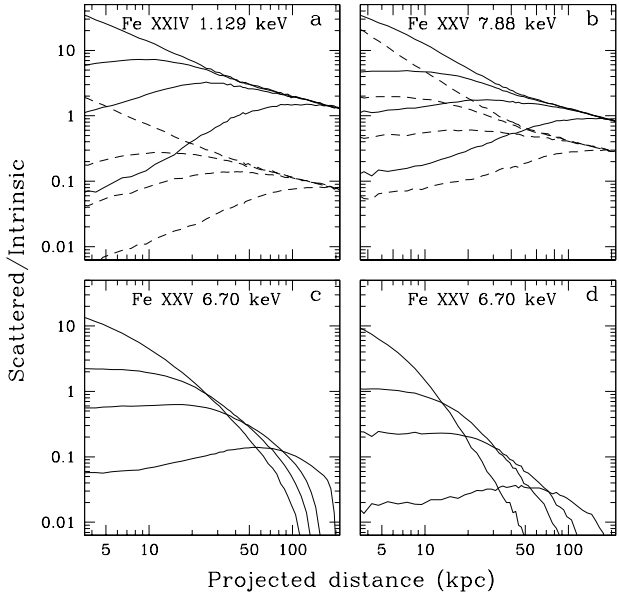}
\caption{{\bf a)} Ratio of the surface brightness of scattered AGN radiation to
the intrinsic brightness of the hot gas in the iron 1.13~keV resonance
line, in the switch-off scenario for M87, at different times after 
AGN switch-off as a function of projected radius. The AGN 
emitted $L_X=0.01\ledd$ in the past. The different solid lines
correspond, from top to bottom, to elapsed times of 5, 50, 100 and 250
thousand years. The dashed lines are the corresponding profiles
for the spectral continuum at the energy of the resonance line. {\bf b)}
Same as Fig.~\ref{m87_lines}a, but for the iron Fe 7.88~keV line. 
{\bf c)} Same as (a), but for the iron 6.70~keV line and a
$10^5$-year outburst ($L_X=0.01\ledd$). {\bf d)} Same as (c), but for a
$2\times 10^4$-year outburst.}   
\label{m87_misc}
\end{figure}

We first consider a scenario in which M87 was bright 
in the past for a long time (at least $\sim$~a million years) until it suddenly
turned off (actually switched to its present low-luminosity state) at
$t_\mathrm{off}=0$. We have plotted in Figs.~\ref{m87_misc}a and
\ref{m87_misc}b for two resonance lines the computed contribution of
scattered radiation from the AGN to the surface brightness as a
function of projected radius as would be measured by observers living
in various epochs $t>0$ after the switch-off. Also shown are the 
corresponding profiles for the continuum emission near the lines. The
relative contribution of scattered AGN emission is a factor of $R\sim
6$ (with a scatter of $\sim 3$ between the lines) larger for the lines
than for the continuum, in agreement with the result of
\S\ref{lines_cont}. We adopted $L_X=0.01\ledd$ for our
examples. Since the discussed effect is proportional to $L_X$, the
results can be easily recomputed for any AGN luminosity.

\begin{figure}
\centering
\includegraphics[width=\columnwidth]{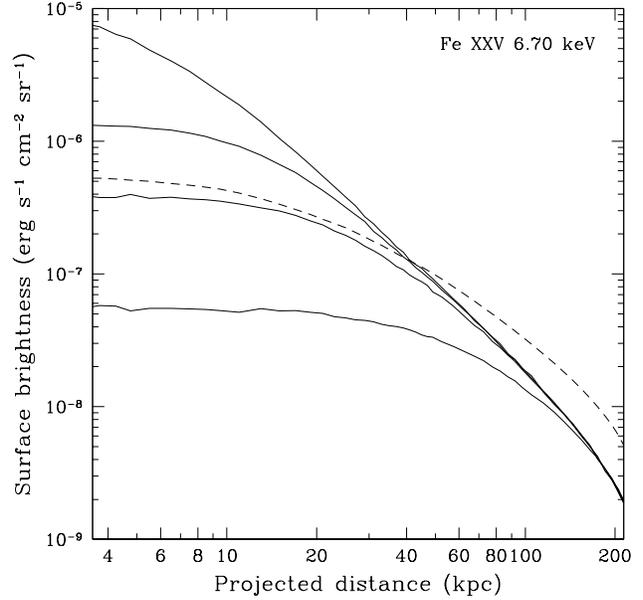}
\caption{M87/Virgo surface brightness of scattered AGN radiation (solid
lines) in the iron 6.70~keV resonance line at different times (as in
Fig.~\ref{m87_misc}) after the AGN switch-off. The dashed line shows
the intrinsic surface brightness profile of the hot gas in the
6.70~keV line, which also is distorted by resonant scattering. The AGN
emitted $L_X=0.01\ledd$ in the past.}
\label{m87_6.70_bright}
\end{figure}

Fig.~\ref{m87_6.70_bright} compares the surface brightness profiles of
scattered AGN radiation and thermal emission in the iron
6.7~keV line, which has the largest equivalent width in our sample. 
It is essential that photon statistics is not expected to
impose strong restrictions on the prospects of detecting scattered AGN
radiation from the Virgo cluster out to $\sim 1$~Mpc from M87 
with the next generation of X-ray spectrometers. 

\begin{figure}
\centering
\includegraphics[width=\columnwidth]{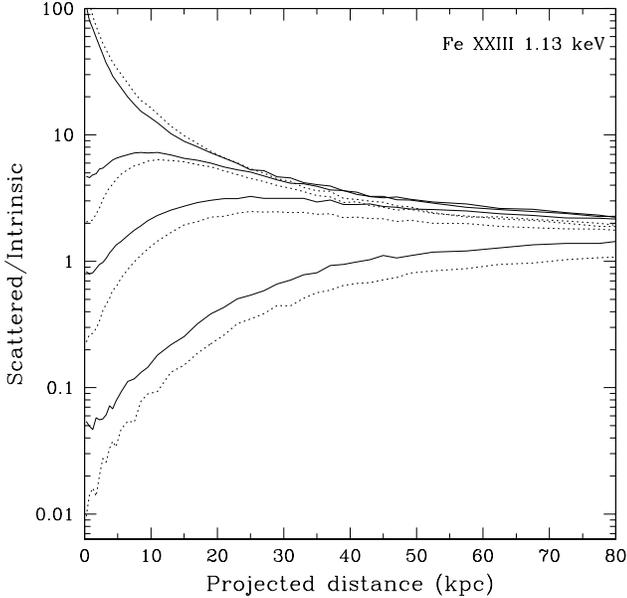}
\caption{Ratio of the surface brightness of scattered AGN radiation to
the intrinsic brightness of the hot gas in the iron 1.13~keV resonance
line, in the switch-off scenario for M87, as a function of projected
radius. The sampled times are as in Fig.~\ref{m87_misc}. The solid lines
correspond to the case of negligible turbulence, and the dotted lines
to $M=0.5$. Note the linear scaling chosen for the projected radius,
as opposed to the logarithmic scaling in Fig.~\ref{m87_misc}.}
\label{m87_1.129_m}
\end{figure}

The finite depth effects that were discussed in \S\ref{depth} play in
the M87 case only a minor role for the X-ray lines, because for them
$\tau_0< 3$ (see Table~\ref{m87_lines}). Fig.~\ref{m87_1.129_m}  
compares two cases, $M=0$ (no turbulence) and $M=0.5$ (significant
turbulence), for the iron 1.13~keV line. The increased turbulence 
leads to a decrease from $\tau_0=3.2$ to $\tau_0=0.7$. We see that  
the relative contribution of scattered AGN emission to the surface
brightness in the central $\sim 10$~kpc is several times larger in the
former (optically-thick) case, except for the early moments. This
increase is mainly due to the delayed escape of photons.  

\begin{figure}
\centering
\includegraphics[width=\columnwidth]{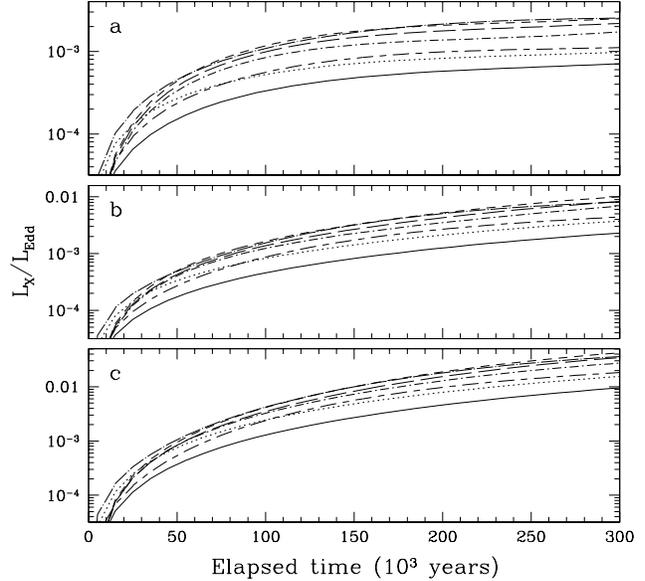}
\caption{{\bf a)} Minimum detectable (for $\delta=10$\% -- see text) past X-ray
luminosity as a function of elapsed time in the switch-off scenario
for M87. The different curves correspond to mapping M87/Virgo in  
different lines: Fe 1.13~keV line (solid), Fe 1.17~keV line (dotted),
Si 1.86~keV line (short-dashed), S 2.46~keV line (long-dashed), Ar 3.14~keV
line (dot-short-dashed), Fe 6.70~keV line (dot-long-dashed) and Fe
7.88~keV line (short-dash-long-dashed). {\bf b)} Same as (a), but for
a $10^5$-year flare. {\bf c)} Same as (b), but for a $2\times
10^4$-year flare.}
\label{m87_detect}
\end{figure}

We can now address the principal question of our study: what
constraints can X-ray observations place on the past luminosity of the M87
central source? Suppose that it is possible in an experiment to
map the distribution of the equivalent width of a particular line over
M87/Virgo with $\delta=10$\% accuracy, which we consider a realistic
value (see the discussion in \S\ref{disc}). Should then the relative
contribution of scattered AGN emission exceed $\delta$ at least at
some projected radius (i.e. at least at the maximum of $EW/EW_0$ 
in Fig.~\ref{m87_misc}), it will be possible to estimate the past AGN
luminosity. The resulting dependence of the minimum detectable
past luminosity of M87 on time after the switch-off is plotted
in Fig.~\ref{m87_detect}a.

We see that in order to have a 10\% contribution of
scattered radiation from the AGN whose luminosity in the past was
$L_X=10^{-4}\ledd= 4\times 10^{43}$~erg s$^{-1}$ (i.e. several hundred
times higher than the combined X-ray luminosity of the nucleus and
jet now), we should be not later than $\sim 3\times 10^4$~years
after the switch-off and look with X-ray telescopes into the M87 core
region ($\rho\lesssim 10$~kpc $\sim 2$~arcmin in the plane of
the sky), as is clear from Fig.~\ref{m87_misc}. If
the luminosity in the past was higher, $L_X=10^{-3}\ledd= 4\times
10^{44}$~erg s$^{-1}$, we must live not later than $\sim 2\times
10^5$~years after the switch-off and look further out (at $\rho\sim
50-200$~kpc $\sim 20$~arcmin from the AGN). The nearly flat ($\propto
t^{-0.6}$) shape of the curves in Fig.~\ref{m87_detect}a, which have
been truncated at $t=3\times 10^5$~years to mimimize the effects of
the finite size of our model gas cloud, suggests that if the X-ray
luminosity of M87 was $\sim$~a few $10^{-3}\ledd\sim
10^{45}$~erg s$^{-1}$ in the past, the maximum allowed elapsed time
would only be restricted by the extent of the gas in the Virgo cluster
($\sim 1$~Mpc), the effective area of the detector and the field of
view of the telescope.

\begin{figure}
\centering
\includegraphics[width=\columnwidth]{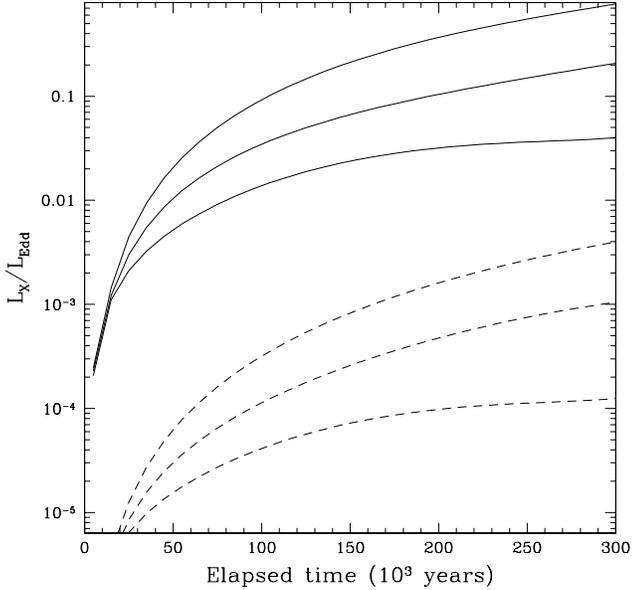}
\caption{Upper limits (for $\delta=100$\%) on the past
X-ray luminosity of the M87 AGN as a function of elapsed time, derived
from XMM-Newton observations of the M87/Virgo central region in the
X-ray spectral continuum below 7~keV, for three cases: switch-off, a
$10^5$-year flare and a $2\times 10^4$-year flare (solid lines from
bottom to top). For comparison shown (dashed lines) are the limits
obtainable if it were possible to map M87/Virgo at $E=20$~keV.} 
\label{m87_detect_cont}
\end{figure}

We may therefore conclude that it should be possible with
future X-ray spectrometers to measure or obtain fairly tight limits on the
past X-ray luminosity of the central source in M87. For comparison, we
can use the available data of XMM-Newton observations of the M87/Virgo
core to derive some weaker limits already now. B\"{o}hringer et
al. (\cite{bohetal01}) have analyzed X-ray spectra taken from a
sequence of rings centered on the M87 nucleus. The most distant ring
has an outer radius $r_\mathrm{out}\approx 60$~kpc, and the innermost
circle has $r_\mathrm{in}\approx 1.25$~kpc. Except for the spectrum of the
central circle to which the direct emission from the AGN contributes,
the individual spectra correspond to optically thin thermal emission
from the gas. The quality of the spectra is apparently good up to
$\sim 7$~keV. We can therefore suggest that the surface brightness of
scattered AGN radiation (having a power-law spectrum) does not exceed
the actually measured surface brightness in the X-ray continuum at
7~keV nowhere between $\rho=1.25$~kpc and 60~kpc. From this
condition we can find an upper limit on the past X-ray luminosity of
M87 as a function of time after the switch-off, similarly as we
derived the limits presented in Fig.~\ref{m87_detect}a based on the
resonance line computations. However, in the previous case
we did not impose any limitations on the projected radius and we
adopted a value $\delta=10$\%, rather than $\delta=100$\% to be used
now. The resulting dependence is shown  in Fig.~\ref{m87_detect_cont}.

We see that the current limits based on the spectral continuum data
are rather weak. For example, the average X-ray
luminosity of the M87 AGN over a long period of $\gtrsim
10^5$~years ending $5\times 10^4$~years ago was not higher than
$5\times 10^{-3}\ledd=2\times 10^{45}$~erg~s$^{-1}$. By comparing
Fig.~\ref{m87_detect_cont} with Fig.~\ref{m87_detect}a, we see that the
current limits could be improved by at least an order of magnitude by
means of fine X-ray spectroscopy. Alternatively, much tighter limits could be
obtained by measuring the surface brightness of M87/Virgo in the
hard X-ray continuum. To illustrate this possibility, we have
repeated our analysis for $E=20$~keV -- see the result in
Fig.~\ref{m87_detect_cont}. Going from $E=7$~keV to 20~keV (again
assuming a $\delta=100$\% scattered contribution relative to the
bremsstralung emission) leads to a factor of $\sim 200$ (!) stronger limits.

\subsubsection{Flare scenario}
\label{m87_flare}

Suppose now that some time ago M87 experienced an outburst that 
was short compared to the characteristic light crossing time of
the gas cloud, i.e. $\Delta\ll 10^6$~years. Some analytic estimates for
this case were obtained in \S\ref{flare}. Figs.~\ref{m87_misc}c and
\ref{m87_misc}d show the computed time evolution of the
scattered/thermal brightness ratio profile for the Fe XXV K$\alpha$
line for outburst durations $\Delta=10^5$~years and $\Delta=2\times
10^4$~years, respectively, and an AGN luminosity $L_X=0.01\ledd$.
Note that for $t\gg\Delta$ the scattered surface
brightness profile depends on the product $L_X\Delta$, rather
than on $L_X$ and $\Delta$ separately. 

Figs.~\ref{m87_detect}b and \ref{m87_detect}c show 
the minimum detectable (for 10\% scattered contribution) AGN outburst
luminosity as a function of elapsed time for the two considered
scenarios. These dependences should be compared with
Fig.~\ref{m87_detect}a representing the switch-off scenario. We see
that until a certain moment $t_\mathrm{crit}\sim\Delta$ the 
minimum detectable luminosity is only a little smaller in the flare
case than in the switch-off case. This is expected, because the
characteristic line-of-sight depth of the illuminated volume at
$t=t_\mathrm{crit}$ is about the same, $\delta\rho\sim
ct_\mathrm{crit}$, in both cases. Only at 
$t\gg\Delta$, does the minimum detectable luminosity become much larger
(by a factor of $\Delta/t$) for short outbursts. This directly follows
from comparison of equations (\ref{ratio_stat}) and (\ref{ratio_short}).

Fig.~\ref{m87_detect_cont} shows upper limits for the same outburst
scenarios but obtained using the XMM-Newton data on the X-ray
continuum emission from M87/Virgo.

\subsection{Cygnus A}
\label{cyga}

\begin{table}
\caption{The brightest resonance X-ray and extreme UV lines of the
Cyg~A intracluster gas.
}
\begin{center}
\begin{tabular}{lccc}\hline\hline
Ion & Energy & Equivalent width & Optical depth \\ 
    & (keV) &(eV)               &               \\
\hline
Fe XXIV  & 0.049 & 1   & 0.9 \\
Fe XXIV  & 0.065 & 2   & 1.8 \\
Fe XXV   & 6.70  & 250 & 1.8 \\
Fe XXV   & 7.88  &  40 & 0.3 \\
Fe XXVI  & 6.97  & 200 & 0.3 \\
\hline
\end{tabular}
\end{center}
\label{cyga_lines}
\end{table}

Our second example is Cyg~A. This is a well-known nearby
($z=0.0562$, i.e. about 20 times more distant than M87 -- 1~arcmin
approximately corresponds to 100~kpc) powerful radio galaxy. Cyg~A has
recently been observed with the Chandra satellite, and detailed information
was obtained on the morphology of X-ray emission from the galaxy
(Young et al. \cite{youetal02}) as well on the intracluster gas surrounding it
(Smith et al. \cite{smietal02}).

In the Chandra and quasi-simultaneous RXTE observations (Young et
al. \cite{youetal02}), hard (up to 100~keV), spatially unresolved
X-ray emission was detected from the position of the radio and
infrared nucleus of Cyg~A. The energy spectrum of this radiation is
power law with photon index $\gamma= 1.5$, heavily absorbed below a
few~keV. The inferred unabsorbed 1--10~keV luminosity of the nucleus
$L_X\sim 5\times 10^{44}$~erg s$^{-1}$. This is four orders of
magnitude more than the X-ray luminosity of the M87 nucleus.

The Chandra observations clearly reveal an intracluster medium. The gas
has complex structure within the central $\sim 100$~kpc,
which apparently is the result of an interaction with the relativistic material
produced by AGN activity. In contrast, the morphology of the gas
is simple at larger radii ($\gtrsim 100$~kpc), namely the gas appears to
be spherical within at least 700~kpc of the nucleus, and
nearly isothermal with $kT\sim 7-8$~keV. The total luminosity of the
intracluster gas is $\sim 10^{45}$~erg s$^{-1}$, which is
comparable to the X-ray luminosity of the central AGN and is a typical
value for rich clusters of galaxies. 

Given the above observational facts, we can estimate the 
contribution of scattered emission from the AGN to the X-ray surface
brightness outside the central 100~kpc of the Cyg~A cluster. Due to
the highly irregular distribution of gas in the innermost region of Cyg~A, our
simulations will not be aimed at this zone. Based on these
considerations and the results of Smith et al. (\cite{smietal02}), we model the
radial distribution of gas density by a beta-model with
$n_0=0.05$~cm$^{-3}$, $\rc=30$~kpc and $\beta=0.5$. The temperature is
assumed to be constant, $kT=5$~keV, within the central $50$~kpc and
also at $r>150$~kpc, $kT=7.5$~keV. The abundance of iron is taken to be 0.35
solar. Our model cloud of gas has an outer boundary at $r=1$~Mpc. 

\begin{figure}
\centering
\includegraphics[width=\columnwidth]{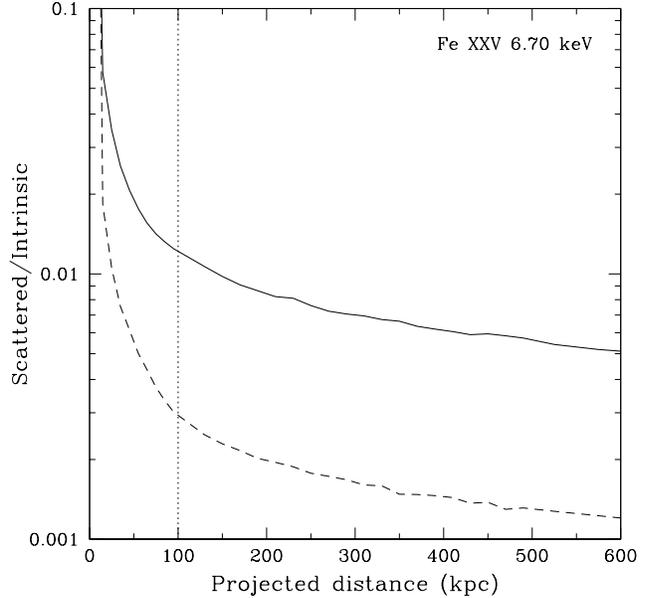}
\caption{Ratio of the surface brightness of scattered AGN radiation to
the intrinsic brightness of the hot gas in the iron 6.70~keV resonance
line, in the stationary scenario for Cyg~A, as a function of
projected radius (the solid line). The AGN X-ray luminosity is assumed
to be equal to its present-day value, $L_X=5\times 10^{44}$~erg
s$^{-1}$. The dashed line shows the corresponding profile for the
spectral continuum at the resonance energy. The vertical
dotted line indicates that the results to the left of it are not reliable.}
\label{cyga_6.70_stat}
\end{figure}

Table~\ref{cyga_lines} lists the strongest emission lines of the
Cyg~A cluster, as implied by our model. The Chandra and RXTE data do
reveal two strong lines,  one near 6.7~keV (Fe K$\alpha$) and another
near 7.9~keV (Fe K$\beta$ plus possibly Ni K$\alpha$). It appears (see
Fig.~7 in Smith et al. \cite{smietal02}) that the measured (with
moderate energy resolution) spectra also do not contradict the
presence of a strong line at the position of Fe Ly$\alpha$ (6.97~keV).

\begin{figure}
\centering
\includegraphics[width=\columnwidth]{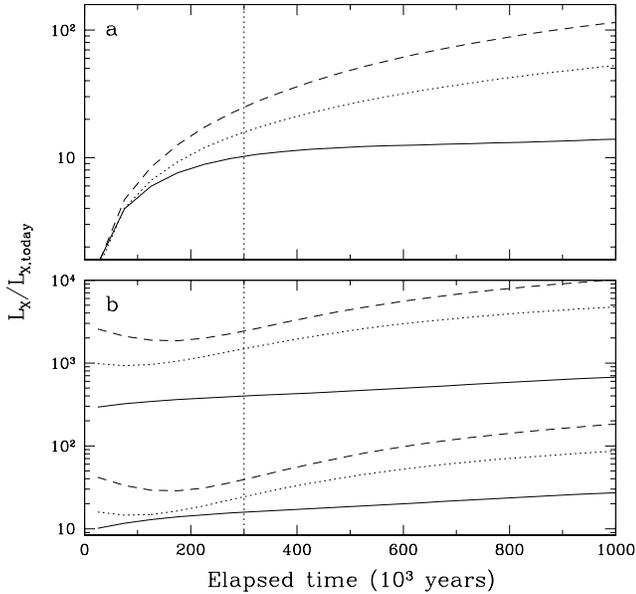}
\caption{{\bf a)} Minimum detectable (for $\delta=10$\% -- see text) past X-ray
luminosity of Cyg~A as a function of elapsed time. The surface
brightness is measured in the iron 6.70~keV line. The solid line
corresponds to the switch-off case ($\Delta\rightarrow\infty$). The
dotted and dashed lines correspond to outbursts of duration $\Delta=5\times
10^5$~years and $\Delta=2\times 10^5$~years, respectively. The
luminosity is normalized to the present-day X-ray luminosity of the
Cyg~A nucleus. The vertical dotted line indicates that the results
to the left of it are not reliable. {\bf b)} Same as (a), but derived from
Chandra observations of the Cyg~A cluster in the X-ray spectral
continuum below 8~keV (for $\delta=100$\%). For comparison shown (the
lower set of lines) are the limits that could be obtained if
it were possible to map the Cyg~A cluster at $E=40$~keV.}
\label{cyga_detect}
\end{figure}

We can estimate from equation (\ref{ratio_stat}) the stationary
contribution of scattered AGN radiation to the surface brightness of
the Cyg~A cluster in the X-ray continuum:
\beqa
\frac{B_{E,\mathrm{cont}}^\mathrm{scat}}{B_{E,\mathrm{cont}}} &=& 
2\times 10^{-3}
\frac{L_X}{5\times 10^{44}\,\mathrm{erg s}^{-1}}
E^{-0.1}\exp(E/7.5\,\mathrm{keV})
\nonumber\\
&&\times\left(\frac{\rho}{30\,\mathrm{kpc}}\right)^{-0.5},
\label{ratio_cyg}
\eeqa
where we have normalized the X-ray luminosity of the Cyg~A nucleus
to its present-day value and used $\gamma=1.5$. 

Fig.~\ref{cyga_6.70_stat} shows a computed Cyg~A radial profile of
the scattered/thermal surface brightness ratio for the iron 6.7~keV
line, assuming that the AGN X-ray luminosity has remained for a
few million years the same as it is now (the stationary scenario). We
see that scattered AGN emission contributes of the order of 1\% at
$\rho\sim 200$~kpc. It then follows that the Cyg~A luminosity should
have been an order of magnitude higher in the past (i.e. $L_X\sim
5\times 10^{45}$~erg s$^{-1}$) for scattered AGN radiation to
contribute $\sim 10$\% to the strong X-ray lines observed from the
intracluster gas (see Fig.~\ref{cyga_detect}).

We can also use the available Chandra data on the X-ray continuum emission from
the Cyg~A cluster to find upper limits on the past X-ray
luminosity of its central source. We adopt the following parameters
for this analysis: $E=8$~keV -- the maximum photon energy at which
thermal bremsstrahlung emission is still detected, and
$\rho_\mathrm{min}=10^2$~kpc and $\rho_\mathrm{max}=6\times 10^2$~kpc
-- the boundaries of the region that is well explored with
Chandra. We plot the resulting upper limits in
Fig.~\ref{cyga_detect}b. We see that the current constraints are very
weak because of the high temperature ($\sim 7$~keV) of the
intracluster gas. For comparison (see Fig.~\ref{cyga_detect}b), 
arcminute-resolution observations of the Cyg~A cluster at $E\sim 40$~keV
could provide limits on its past luminosity similar to those 
obtainable by mapping the cluster in resonance lines (see
Fig.~\ref{cyga_detect}a).

\section{Discussion and conclusions}
\label{disc}

The main results of this paper are as follows.
\begin{itemize}

\item
We have demonstrated that there are two viable observational
strategies for constraining the past X-ray luminosity of galactic
nuclei with the next generation of X-ray telescopes. One is to search
for scattered AGN radiation in the spectral continuum 
at high energies $E\gg kT(1+z)$, and the other is to perform imaging 
in bright resonance X-ray lines. The first approach appears to be
particularly promising for studying distant (at $z\gtrsim 0.5$)
powerful quasars and their environments (\S\ref{cont}). 

\item
We have shown that the relative contribution of
scattered radiation is typically larger by a significant factor of
3--10 in a resonance line than in the neighboring continuum
(\S\ref{lines_cont}).

\item
We have estimated the level of constraints that could be derived 
from future observations on the past X-ray luminosity of the nearby
M87 and Cyg~A active galaxies (\S\ref{num_sim}). The upper limits imposed
by the available XMM-Newton and Chandra X-ray continuum data are typically
1--2 orders of magnitude weaker. 
\end{itemize}

Although we have been mostly discussing clusters of galaxies and
their central dominant galaxies, the same method
can also be applied to groups of galaxies and isolated giant elliptical
galaxies, which also are large reservoirs of hot ionized gas capable
of scattering AGN radiation. The main advantage of clusters 
is the larger extent of intracluster gas, which enables studying
previous galactic activity on longer timescales (up to a few times
$10^6$~years for the richest clusters) than in the case of groups of
galaxies and early-type galaxies (up to a few times $10^5$~years). 
On the other hand, the latter two classes of object have their own
advantage that typical gas temperatures for them are $kT\sim
1$~keV and the factor $\exp(E/kT)$ becomes significant already in the
standard X-ray band ($< 10$~keV). In particular, one can use the K$\alpha$ line
of He-like iron at 6.7~keV. Indeed, the corresponding ions
(Fe XXV) are still abundant at such low temperatures, but collisional
excitation of the ions leading to emission in the line is already
inefficient.

Although we have only considered the case where the AGN is an
isotropic source, in reality angular anisotropy at some level is
expected. Therefore, the contribution of scattered emission from
the AGN to the surface brightness may be larger or smaller in some
regions than predicted assuming source isotropy. We refer the reader to the
papers by Gilfanov et al. (\cite{giletal87a}) and Wise \& Sarazin
(\cite{wissar92}) for a detailed discussion of scattered surface
brightness profiles that may arise in the case of beamed source emission.

In the discussion of our simulations for the M87 and Cyg~A galaxies
and their associated clusters of galaxies we used a fiducial value of
10\% as the minimum detectable contribution of scattered AGN emission
to a resonance line. We realize that significant efforts will be
required to achieve this or better detection level experimentally. One
difficulty is that the measured surface brightness is the integral of
emission (thermal plus scattered) along a given line of sight, and
there may be present gas with different temperatures and element
abundances along this line of sight. Both types of variations will
have an effect on the equivalent line width. In principle it should be
possible to determine and subtract the underlying temperature and
abundance radial trends from a careful analysis of the
spectral-imaging data. The remaining uncertainties are expected to be
largest for cluster central regions, where both the gas temperature
and element abundances vary significantly, for example in the M87
case. On the other hand, this problem is not expected to be severe
outside cluster cores. For example, in the Cyg~A cluster the gas
appears to be nearly isotropic outside the central $\sim 200$~kpc (see
Smith et al. \cite{smietal02}), whereas the equivalent width of the
6.7~keV iron emission line changes by only $\sim 10$\% when the gas temperature
varies by $0.5$~keV around $kT\sim 7.5$~keV. Therefore, achieving the
10\% detection level appears realistic.

It is clear that the potential of the method can be fully realized
only when it is possible to resolve the interesting resonance lines from
neighbouring lines. Indeed, lines other than resonance lines can
contribute significantly to the intrinsic emission of the gas, but
little to the scattered AGN emission. Moreover,
it is desirable that all important lines in blends be resolved, such
as the permitted, intercombination and forbidden lines in the 
complex around the 6.7~keV (Fe XXV K$\alpha$) line. Even a small contribution
of scattered emission from the AGN could then be made manifest by
comparing the surface brightness profiles measured in these lines. We
are therefore looking forward to future high-energy astrophysics 
missions such as Astro-E2, Constellation-X and XEUS that will provide
the required energy resolution ($\sim$~a few eV). 

Finally, we would like to note that the constraints on the past AGN
X-ray luminosity could be further improved  by means of X-ray 
polarimetry, especially if it became possible to measure polarization in X-ray
resonance lines. The radiation of a central source gets strongly
polarized upon scattering in a beta-cluster (with a typical resulting degree of
polarization $P\sim 60$\%, Sunyaev 1982), whereas the intrinsic
emission of the intracluster gas in X-ray resonance lines is also
polarized but to a much lesser degree ($P\lesssim 10$\%, Sazonov et
al. 2002). Thus, even a few per cent contribution of scattered AGN
radiation to the surface brightness in a resonance line would be
manifest in polarimetric observations.

\acknowledgements
We thank the referee, Luca Ciotti, for comments
that helped to improve the presentation of the paper. SS acknowledges support
from a Peter Gruber Foundation Fellowship. RS as a Gordon Moore
Scholar thanks Caltech for hospitality during the completion of this
paper. This research was partially supported by the Russian Foundation
for Basic Research (projects 00-02-16681 and 00-15-96649) and by the
program of the  Russian Academy of Sciences "Astronomy (Nonstationary
astronomical objects)".


\begin{thebibliography}{}

\bibitem[1973]{aldpeq73} Aldrovandi, S.M.V., \& Pequignot, D. 1973,
A\&A, 25, 137
 
\bibitem[1989]{andgre89} Anders, E., \& Grevesse, N. 1989, Geochimica
et Cosmochimica Acta, 53, 197
 
\bibitem[1985]{arnrot85} Arnaud, M., \& Rothenflug, R. 1985, A\&AS, 60, 425

\bibitem[1996]{arnaud96} Arnaud, K.A. 1996, Astronomical Data
Analysis Software and Systems V, eds. Jacoby G. and Barnes J., ASP
Conf. Series volume 101, 17

\bibitem[1997]{bahetal97} Bahcall, J.N., Kirhakos, S., Saxe, D.H., \&
Schneider, D.P. 1997, ApJ, 479, 642
  
\bibitem[1995]{bintab95} Binney, J., \& Tabor, G. 1995, MNRAS, 276, 663 

\bibitem[1991]{biretal91} Biretta, J.A., Stern, C.P., \& Harris,
D.E. 1991, AJ, 101, 1632

\bibitem[2001]{bohetal01} B\"ohringer, H., Belsole, E., Kennea, J.,
Matsushita, K., Molendi, S., Worrall, D.~M., Mushotzky, R.F.,
Ehle, M., Guainazzi, M., Sakelliou, I., Stewart, G., Vestrand,
W.T., \& Dos  Santos, S. 2001, A\&A, 365, 181

\bibitem[1997]{boyetal97} Boyce, P.J., Disney, M.J., Blades, J.C.,
Boksenberg, A., Crane, P., Deharveng, J.M., Macchetto, F.D., Mackay, C.D.,
\& Sparks, W.B. 1997, MNRAS, 298, 121

\bibitem[1998]{buretal98} Burderi, L., King, A.R., \& Szuszkiewicz
1998, ApJ, 509, 85

\bibitem[2002]{caretal02} Carilly, C.L., Harris, D.E., Pentericci, L., et
al. 2002, ApJ, 567, 781

\bibitem[1976]{cavfus76} Cavaliere, A. \& Fusco-Femiano, R. 1976,
A\&A, 49, 137

\bibitem[1950]{chandra50} Chandrasekhar, S. 1950, Radiative Transfer,
Oxford, Clarendon Press

\bibitem[1997]{cioost97} Ciotti, L., \& Ostriker, J.P. 1997, ApJ, 487,
L105

\bibitem[2001]{cioost01} Ciotti, L., \& Ostriker, J.P. 2001, ApJ, 551,
131

\bibitem[2002]{crasun02} Cramphorn, C.K., \& Sunyaev, R.A. 2002, A\&A
(in press)

\bibitem[1999]{craetal99} Crawford, C.S., Lehmann, I., Fabian, A.C.,
Bremer, M.N., \& Hasinger, G. 1999, MNRAS, 308, 1159

\bibitem[2001]{fabetal01} Fabian, A.C., Crawford, C.S., Ettori, S., \&
Sanders, J.S. 2001, MNRAS, 322, L11

\bibitem[2002]{fanetal02} Fang, T.D., Davis, D.S., Lee, J.C.,
Marshall, H.L., Bryan, G.L., \& Canizares, C.R. 2002, ApJ, 565, 86 

\bibitem[2002]{finetal02} Finoguenov, A., Matsushita, K., B\"ohringer,
H., Ikebe, Y., Arnaud, M. 2002, A\&A, 381, 21 

\bibitem[1987a]{giletal87a} Gilfanov, M.R., Sunyaev, R.A., \& Churazov,
E.M. 1987a, Soviet Astron. Lett., 13, 233 

\bibitem[1987b]{giletal87b} Gilfanov, M.R., Sunyaev, R.A., \& Churazov, 
E.M. 1987b, Soviet Astron. Lett., 13, 3

\bibitem[1999]{harwor99} Hardcastle, M.J., \& Worrall, D.M. 1999, MNRAS,
309, 969

\bibitem[2000]{haretal00} Harris, D.E., Nulsen, P.E.J., Ponman, T.J.,
Bautz, M., Cameron, R.A., David, L.P., Donnelly, R.H., Forman, W.R.,
Grego, L., Hardcastle, M.J., Henry, J.P., Jones, C., Leahy, J.P.,
Markevitch, M., Martel, A.R., McNamara, B.R., Mazzotta, P., Tucker,
W., Virani, S.N., \& Vrtilek, J. 2000, ApJ, 530, L81

\bibitem[2001]{haretal01} Hardcastle, M.J., Birkinshaw M. \& Worrall
D.M. 2001, ApJ, 323, L17

\bibitem[1989]{hernquist89} Hernquist, L. 1989, Nature, 340, 687

\bibitem[1997]{hooetal97} Hooper, E.J., Impey, C.D., \& Foltz,
C.B. 1997, ApJ, 480, L95

\bibitem[1992]{kaastra92} Kaastra, J.S. 1992, An X-Ray Spectral Code
for Optically Thin Plasmas (Internal SRON-Leiden Report, updated
version 2.0)

\bibitem[1996]{koyetal96} Koyama, K., Maeda, Y., Sonobe, T.,
Takeshima, T. Tanaka, Y., \& Yamauchi, S. 1996, PASJ, 48, 249

\bibitem[1986]{linshi86} Lin, D.N.C., \& Shields, G.A. 1986, ApJ, 305,
28

\bibitem[1997]{macetal97} Macchetto, F.D., Marconi, A., Axon, D.J.,
Capetti, A., Sparks, W.B., \& Crane, P. 1997, ApJ, 489, 579

\bibitem[1997]{maretal97} Marconi, A., Axon, D.J., Macchetto, F.D., 
Capetti, A., Sparks, W.B., \& Crane, P. 1997, MNRAS, 289, L21

\bibitem[2002]{matetal02} Matsushita, K., Belsole, E., Finoguenov, A.,
\& Boehringer, H. 2002, A\&A 386, 77

\bibitem[1999]{mcletal99} McLure, R.J., Kukula, M.J., Dunlop, J.S., Baum,
S.A., O'Dea, C.P., \& Hughes, D.H. 1999, MNRAS, 308, 377

\bibitem[2001]{mcldun01} McLure, R.J., \& Dunlop, J.S. 2001, MNRAS, 321, 515

\bibitem[1985]{mewetal85} Mewe, R., Gronenschild, E.H.B.M.\& van der
Oord, G.H.J. 1985, A\&AS, 62, 197

\bibitem[1990]{minshi90} Mineshige, S., \& Shields, G.A. 1990, ApJ,
351, 47

\bibitem[2001]{muretal01} Murakami, H., Koyama, K., \& Maeda, Y. 2001,
ApJ 558, 687

\bibitem[1993]{murche93} Murphy, B.W., \& Chernoff, D.F. 1993, ApJ, 418, 60

\bibitem[2000]{oweetal00} Owen, F.N., Eilek, J.A., \& Kassim, N.E. 2000,
ApJ, 362, 449

\bibitem[1988]{rees88} Rees, M.J. 1988, Nature, 333, 523

\bibitem[1993]{sarwis93} Sarazin, C.L., \& Wise, M.W. 1993, ApJ, 411, 55

\bibitem[2002]{sazetal02} Sazonov, S.Yu., Churazov, E.M., \& Sunyaev,
R.A. 2002, MNRAS, 333, 191
 
\bibitem[2000]{schetal00} Schade, D.J., Boyle, B.J., \& Letawsky, M. 2000,
MNRAS, 315, 498 

\bibitem[2001]{shietal01} Shibata, R., Matsushita, K., Yamasaki, N.Y.,
Ohashi, T., Ishida, M., Kikuchi, K., Boehringer, H., \& Matsumoto,
H. 2001, ApJ, 549, 228

\bibitem[1980]{shoyas90} Sholomitskii, G.B. \& Yaskovich, A.L. 1990,
Soviet Astron. Lett., 16, 383

\bibitem[1982]{shuste82} Shull, J.M., \& van Steenberg, M. 1982, ApJS, 48, 95 

\bibitem[1996]{sieetal96} Siemiginowska, A., Czerny, B., \& Kostyunin,
V. 1996, ApJ, 458, 491

\bibitem[2002]{smietal02} Smith, D.A., Wilson, A.S., Arnaud, K.A.,
Terashima, Y., \& Young, A.J. 2002, ApJ, 565, 195

\bibitem[1982]{sunyaev82} Sunyaev, R.A. 1982, Soviet Astron. Lett., 8, 175

\bibitem[1993]{sunetal93} Sunyaev, R.A., Markevitch, M., \& Pavlinsky
M. 1993, ApJ, 407, 606

\bibitem[1998]{sunchu98} Sunyaev, R.A., \& Churazov, E.M. 1998, MNRAS,
297, 1279

\bibitem[1992]{sunyaev92} Sunyaev, R.A. 1982, Astron. Lett., 8 ,175

\bibitem[1996]{verfer96} Verner, D.A., \& Ferland, G.J. 1996, ApJS, 103, 467

\bibitem[1996]{veretal96} Verner, D.A., Verner, E.M., \& Ferland,
G.J. 1996, Atomic Data and Nuclear Data Tables, 64, 1

\bibitem[1997]{voronov97} Voronov, G.S. 1997, Atomic Data and Nuclear
Data Tables, 65, 1

\bibitem[2002]{wilyan02} Wilson, A.S., \& Yang, Y. 2002, ApJ, 568, 133

\bibitem[1990]{wissar90} Wise, M.W., \& Sarazin, C.L. 1990, ApJ, 363, 344

\bibitem[1992]{wissar92} Wise, M.W., \& Sarazin, C.L. 1992, ApJ, 395, 387

\bibitem[2001]{woretal01} Worrall, D.M., Birkinshaw, M., Hardcastle,
M.J., \& Lawrence, C.R. 2001, MNRAS, 326, 1127

\bibitem[1990]{yaqser00} Yaqoob, T. \& Serlemitsos, P. 2000, ApJ,
544, L95

\bibitem[2002]{youetal02} Young, A.J., Wilson, A.S., Terashima, Y.,
Arnaud, K.A., \& Smith, D.A. 2002, 564, 176

\bibitem[1990]{zombeck90} Zombeck, M.V. 1990, Handbook of Astronomy
and Astrophysics, Second Edition (Cambridge, UK: Cambridge University
Press)

\end{thebibliography}
\end{document}